\documentclass[fleqn,usenatbib]{mnras}

\usepackage{newtxtext,newtxmath}

\usepackage[T1]{fontenc}

\DeclareRobustCommand{\VAN}[3]{#2}
\let\VANthebibliography\thebibliography
\def\thebibliography{\DeclareRobustCommand{\VAN}[3]{##3}\VANthebibliography}

\usepackage{graphicx}	
\usepackage{amsmath}	
\usepackage{ulem}
\usepackage[utf8]{inputenc}
\usepackage{scalerel}
\usepackage{tikz}
\usetikzlibrary{svg.path}

\definecolor{orcidlogocol}{HTML}{A6CE39}
\tikzset{
  orcidlogo/.pic={
    \fill[orcidlogocol] svg{M256,128c0,70.7-57.3,128-128,128C57.3,256,0,198.7,0,128C0,57.3,57.3,0,128,0C198.7,0,256,57.3,256,128z};
    \fill[white] svg{M86.3,186.2H70.9V79.1h15.4v48.4V186.2z}
                 svg{M108.9,79.1h41.6c39.6,0,57,28.3,57,53.6c0,27.5-21.5,53.6-56.8,53.6h-41.8V79.1z M124.3,172.4h24.5c34.9,0,42.9-26.5,42.9-39.7c0-21.5-13.7-39.7-43.7-39.7h-23.7V172.4z}
                 svg{M88.7,56.8c0,5.5-4.5,10.1-10.1,10.1c-5.6,0-10.1-4.6-10.1-10.1c0-5.6,4.5-10.1,10.1-10.1C84.2,46.7,88.7,51.3,88.7,56.8z};
  }
}

\newcommand\orcidicon[1]{\href{https://orcid.org/#1}{\mbox{\scalerel*{
\begin{tikzpicture}[xscale=1,yscale=-1, transform shape]
\pic{orcidlogo};
\end{tikzpicture}
}{|}}}}

\newcommand{\tess}{{\it TESS}}
\newcommand{\asassn}{{\it ASAS-SN}}

\newcommand{\plato}{{\it PLATO}}

\newcommand{\kepler}{{\it Kepler}}
\newcommand{\ktwo}{{\it K2}}

\title[QPOs in Accreting White Dwarfs]{Discovery of Persistent Quasi-Periodic Oscillations in Accreting White Dwarfs: A New Link to X-ray Binaries}

\author[M. Veresvarska et al.]{M. Veresvarska$^{\orcidicon{0000-0002-0146-3096}}$$^{1}$\thanks{E-mail: martina.veresvarska@durham.ac.uk},
S. Scaringi$^{\orcidicon{0000-0001-5387-7189}}$$^{1,2}$,
C. Knigge$^{\orcidicon{0000-0002-1116-2553}}$$^{3}$,
J. Paice$^{\orcidicon{0000-0003-1149-1741}}$$^{1}$,
D.A.H. Buckley$^{\orcidicon{0000-0002-7004-9956}}$$^{4,5,6}$,\newauthor
N. Castro Segura$^{\orcidicon{0000-0002-5870-0443}}$$^{7}$,
D. de Martino$^{\orcidicon{0000-0002-5069-4202}}$$^{2}$,
P. J. Groot$^{\orcidicon{0000-0002-4488-726X}}$$^{8,4,5}$,
A. Ingram$^{\orcidicon{0000-0002-5311-9078}}$$^{9}$,
Z.A. Irving$^{\orcidicon{0009-0006-0951-3429}}$$^{3}$,
P. Szkody$^{\orcidicon{0000-0003-4373-7777}}$$^{10}$
\\
$^{1}$Centre for Extragalactic Astronomy, Department of Physics, Durham University, South Road, Durham, DH1 3LE\\
$^{2}$INAF-Osservatorio Astronomico di Capodimonte, Salita Moiariello 16, I-80131 Naples, Italy\\
$^{3}$School of Physics and Astronomy, University of Southampton, Highfield, Southampton SO17 1BJ, UK\\
$^{4}$South African Astronomical Observatory, PO Box 9, Observatory 7935, Cape Town, South Africa\\
$^{5}$Department of Astronomy, University of Cape Town, Private Bag X3, Rondebosch 7701, South Africa\\
$^{6}$Department of Physics, University of the Free State, P.O. Box 339, Bloemfontein 9300, South Africa\\
$^{7}$Department of Physics, University of Warwick, Gibbet Hill Road, Coventry CV4 7AL, UK\\
$^{8}$Department of Astrophysics/IMAPP, Radboud University, P.O. Box 9010, 6500 GL Nijmegen, The Netherlands\\
$^{9}$School of Mathematics, Statistics and Physics, Newcastle University, Herschel Building, Newcastle upon Tyne, NE1 7RU, UK\\
$^{10}$Department of Astronomy, University of Washington, Seattle, WA 98195, USA\\
}

\date{Accepted XXX. Received YYY; in original form ZZZ}

\pubyear{2023}

\begin{document}
\label{firstpage}
\pagerange{\pageref{firstpage}--\pageref{lastpage}}
\maketitle

\begin{abstract}
Almost all accreting black hole and neutron star X-ray binary systems (XRBs) exhibit prominent brightness variations on a few characteristic time-scales and their harmonics. These quasi-periodic oscillations (QPOs) are thought to be associated with the precession of a warped accretion disc, but the physical mechanism that generates the precessing warp remains uncertain. Relativistic frame dragging (Lense-Thirring precession) is one promising candidate, but a misaligned magnetic field is an alternative, especially for neutron star XRBs. Here, we report the discovery of 5 accreting white dwarf systems (AWDs) that display strong optical QPOs with characteristic frequencies and harmonic structures that suggest they are the counterpart of the QPOs seen in XRBs. Since AWDs are firmly in the classical (non-relativistic) regime, Lense-Thirring precession cannot account for these QPOs. By contrast, a weak magnetic field associated with the white dwarf can drive disc warping and precession in these systems, similar to what has been proposed for neutron star XRBs. Our observations confirm that magnetically driven warping is a viable mechanism for generating QPOs in disc-accreting astrophysical systems, certainly in AWDs and possibly also in (neutron star) XRBs. Additionally, they establish a new way to estimate magnetic field strengths, even in relatively weak-field systems where other methods are not available.

\end{abstract}

\begin{keywords}
accretion -- accretion discs -- cataclysmic variables -- magnetic fields
\end{keywords}



\section{Introduction}
\label{s:Intro}

Accreting white dwarfs (AWDs) are binary systems in which a white dwarf (WD) accretes material from a donor star. The dominant population of AWDs are referred to as cataclysmic variables (CVs) where mass transfer occurs through Roche lobe overflow. In cases where the donor is a degenerate or semi-degenerate star they are referred to as AM CVn systems \citep{Solheim2010}.

The WD magnetic field strength can alter the dynamics of the accretion flows in AWDs. If the WD has a sufficiently strong magnetic field ($\gtrsim$ 10$^{6}$ G) the disc material is disrupted at the magnetospheric radius and is further accreted via the magnetic field lines onto the magnetic poles of the WD. If the magnetic field is strong enough such that the magnetospheric radius lies beyond the disc circularisation radius then no disc can form and accretion proceeds only via accretion streams and columns (both referred to as Intermediate Polars; \citealt{Norton2004,Norton2008ApJ...672..524N}). For systems possessing the strongest WD magnetic fields, the accretion flow only follows the magnetic field lines and the WD spin and system orbital period are synchronised (systems referred to as Polars; \citealt{Cropper1990SSRv...54..195C}).

Quasi-period oscillations (QPOs) are non-coherent brightness variations in the X-ray flux, widely recognised in X-ray binary systems (XRBs). In the context of XRBs the flux fluctuations manifest as characteristically broad features in the power spectra, due to the unstable quasi-periodic nature of the signal. They are present both in neutron star (NS) and black hole (BH) XRBs with a variety of different types (see \citealt{Ingram2019} for a detailed review). BH XRBs exhibit low ($\lesssim$ 30 Hz) and high ($\gtrsim$ 60 Hz) frequency QPOs, with low frequency QPOs being somewhat more common. The low frequency QPOs show different broad types, depending on their strength, width and frequency, one of which is known as Type-C. The broadness of a QPO is characterised by its quality factor $Q = \frac{\nu_{0}}{2 \Delta}$, where $\nu_{0}$ is the centroid frequency of the Lorentzian representing the QPO and $\Delta$ its half width at half maximum. Type-C QPOs in particular are strong and narrow, with $Q \gtrsim 8$, and are also known to display strong harmonics. Similarly, NS XRBs show kHz QPOs and low frequency QPOs, however as opposed to BH XRBs the kHz QPOs are more common. They also display colour evolution on the hardness-intensity diagram, similarly to BH XRBs \citep{Homan2001PhDT.......239H,Belloni2016ASSL..440...61B}. Based on the QPO-colour evolution they are classified as FBO (flaring branch oscillations), NBO (normal branch oscillations) and HBO (horizontal branch oscillations). Type-C and HBO QPOs are known to follow a linear correlation with a broad-band aperiodic low-frequency break \citep{Wijnands1999}, with a characteristic decrease in power with increasing frequency. 

There have been numerous QPOs reported in AWD systems in the literature (see e.g. \citealt{Warner2004PASP..116..115W} for review). The QPOs in the context of AWDs, as opposed to XRBs, usually refer to a transitional and temporary periodic signal in the optical whose amplitude can vary significantly. The first report of a QPO in AWDs was in \citet{Patterson1977ApJ...214..144P}, where a fast ($\sim$50s) oscillation in the light curve was detected in the outburst of RU Peg. Another type of quasi-periodic signals in AWDs are so-called dwarf novae oscillations (DNOs) \citep{Warner1972NPhS..239....2W}, which however appear to display a somewhat more coherent period. One possible explanation is that these are associated to g-mode pulsations of the WD \citep{Warner1998IAUS..185..321W,Woudt2005ASPC..330..325W,Townsley2004ApJ...600..390T,Townsley2016arXiv160102046T}. The DNO periods are usually quite fast ($\sim$10\,s) and follow a relation to the longer QPOs so that $P_{QPO} \approx 16 \times P_{DNO}$. In \citet{Warner2003MNRAS.344.1193W} a further sub-type of DNOs ($P_{DNO} \sim$20s), the long period DNOs (lpDNOs with $P_{lpDNO} \sim$80$-$100s) are also discussed. These are thought to be empirically related to the DNOs and QPOs such that $P_{lpDNO} \approx P_{DNO} \approx \frac{1}{4} P_{QPO}$ and are usually associated to high mass transfer rate systems and are sometimes detected as doubles.

Other types of QPOs in AWDs have been observed, such as the broad feature in the high state of magnetic AWD TX Col \citep{Littlefield2021AJ....162...49L}. The QPO in TX Col as observed by \tess\ spans from $\sim$10 to $\sim$20 d$^{-1}$ and in width resembles the QPO in AM CVn SDSS J1908+3940 reported by \citet{Kupfer2015}. These broad QPOs and their lack of harmonics resembles more a broad-band feature of the PSD most likely associated to mass transfer variation in the accretion disc (see \citealt{Scaringi2014}), rather then a quasi-coherent signal as in XRBs. Similarly to \citet{Warner2003MNRAS.344.1193W}, all of these QPOs have been detected in the optical, with no known X-ray counterparts.
The study by \citet{Warner2003MNRAS.344.1193W} extensively reports QPOs in AWD systems, drawing a comparison between the reported QPOs to those observed in XRBs \citep{Wijnands1999}. WZ Sge, a target in \citet{Warner2003MNRAS.344.1193W} and also studied in this work, was reported to exhibit short-period DNOs at $\sim$27.87 s and 28.95 s and a $\sim$740 s QPO. While the 27.87 s signal has been linked to the spin of WZ Sge \citep{Patterson1980ApJ...241..235P}, the QPO is attributed to the retrograde precession of a geometrically thick disc \citep{Warner2002}. However, the absence of a $\sim$740 s QPO signal in WZ Sge from \tess\ short cadence data and other objects displaying QPOs in \citet{Warner2003MNRAS.344.1193W} suggests these signals are transitional. It is also important to note that the reported correlation between AWDs and XRBs in \citet{Warner2003MNRAS.344.1193W} is not based on broad-band features and QPO frequencies, but rather on transitional QPOs and DNOs, which may represent a separate class of signals. Therefore, the presence of these signals in AWDs could be driven by a different physical process than those observed in XRBs. A further important distinction between all of the above reported QPOs in AWDs and XRBs is, that there have never been reported harmonics of any kind of QPOs in AWDs, as opposed to XRBs.


Here we try to characterise QPOs in AWDs in direct analogy to those in XRBs using self-similar analysis techniques in order to better understand their physical origin. In doing so we report the discovery of 5 AWD systems showing persistent QPOs in optical detected by \tess\, at $\sim$ 1.3 $-$ 3 $\times$ 10$^{-4}$ Hz (1 $-$ 2 hours), which we use to revise the relation to XRBs. We also report on the first instance of harmonics being discovered for QPOs in AWDs in 3 of the 5 reported systems. \tess\ has an archival database of $\sim$1200 AWDs, with even more AWD candidates. After visually inspecting the closest sample of AWDs we recovered the 5 examples reported here. However, it is likely that many more are present in the TESS database and that many more could be uncovered with an instrument with a better signal-to-noise ratio. In Section \ref{s:Obs} the observations used are described. Section \ref{s:Methods} discusses the methods and analysis used as well as the construction of the time-averaged power spectra (TPS) with a similar methodology to what is conventionally employed in analysis of QPOs in XRBs (Sections \ref{ss:PSDs} and \ref{ss:PSDfit}). We also discuss the statistical significance of the observed broad-band frequency features in the PSD used in the analysis (Section \ref{ss:break}). In Section \ref{s:results} we report that the QPOs and broad-band components of the TPS fit from Section \ref{ss:PSDfit} appear to follow the observational correlation between the QPO frequency and the low frequency break from \citet{Wijnands1999}. In Section \ref{s:Discussion} we introduce a proposed model to explain the observed QPOs, based on magnetically driven precession of the inner accretion flow due to the interaction between a weak accretor magnetic field and it's inner accretion flow. We then discuss the implications of the aforementioned results with conclusions drawn in Section \ref{s:Conclusions}.

\section{Observations}
\label{s:Obs}

The data used in this work were obtained by \tess\ and can be accessed on the Mikulski Archive for Space Telescopes (MAST\footnote[1]{\label{mast}\url{https://mast.stsci.edu/portal/Mashup/Clients/Mast/Portal.html}}). \tess\ is a space-based mission that observes the entire sky with 4 cameras and a 24$^{\circ}\times$96$^{\circ}$ continuous field of view. For a selected number of targets it produces photometry with either 2 minutes or 20 second cadence and an overall full frame image of the CCD is available each 30 or 10 minutes. A detailed description of the instrument can be found in the \tess\ Instrument Handbook\footnote[2]{\label{Handbook}\url{https://archive.stsci.edu/files/live/sites/mast/files/home/missions-and-data/active-missions/tess/_documents/TESS_Instrument_Handbook_v0.1.pdf}}. The observing strategies of \tess\ splits the sky into southern and northern sectors, each 24$^{\circ}\times$96$^{\circ}$ in size and observes each sector for 1 month. Including the downlink time and other technical procedures this translates to $\sim$ 27 days worth of scientific data for each sector.

The details of the number of \tess\ sectors available for each object studied here, as well as the chosen cadence, are presented in Table \ref{tab:obs}. The short cadence was used for the objects that had all the available sectors in 20\,s cadence (WZ She and GW Lib). If the short 20~s cadence was only available for some of the sectors, 120~s cadence was used for those particular objects (CP Pup, T Pyx, and V3101 Cyg). 

\tess\ data were downloaded and cosmic rays were removed using the \texttt{Lightkurve} package\footnote[3]{\url{https://docs.lightkurve.org/index.html}}. The Simple Aperture Photometry (SAP) flux is used to retain intrinsic variability of the systems while avoiding the transit detection optimisation of the Pre-search Data Conditioning (PDCSAP) flux. Data points are further excluded if their quality flag $>$ 0.

\tess\ provides photometry in units of e$^{-}$s$^{-1}$, which does not provide any information on the absolute flux of the target, only relative brightness. To obtain absolute photometry in mJy, we attempt to convert these measurements using quasi-simultaneous ground-based observations from \asassn, without accounting for any bolometric correction. This conversion is based on assuming a linear relation between the \tess\ and \asassn\ fluxes. This method has been previously used \citep{Scaringi2022,Veresvarska2024}, but it requires quasi-simultaneous coverage of the target by both \tess\ and \asassn. Such coverage was achieved only for CP Pup, as detailed in \citet{Veresvarska2024}. For the remaining targets no conversion is adopted, due to lack of quasi-simultaneous coverage by \tess\ and \asassn. Whereas this does not pose a problem for the analysis presented in this paper, lack of ground-based calibration prohibits any measurements of the energetics being made. We note that with the advent of more synoptic sky surveys (e.g. BlackGEM \citealt{blackgem}) space-based calibrations will be more easily achieved due to wider sky coverage. 

There is a linear trend in the data of GW Lib and V3101 Cyg, which has been removed using a similar methodology as in \citet{Kupfer2015}. Similar linear trends due to systematic effects have often been found to affect long-term data, especially in the \kepler\ and \ktwo\ mission \citep{Kupfer2015,Solanki2021MNRAS.500.1222S}. These trends can alter the flux levels over time and induce low frequency power excess and hence need to be removed. This is done by fitting a linear trend to the light curve smoothed on a 2~hr timescale which is subsequently subtracted from the original light curve. The correction was done for the whole sector for GW Lib and for each half sector for V3101 Cyg by fitting a straight line to the time-averaged light curve and correcting the non-averaged light curve for the obtained trend. The corrected light curve of CP Pup has already been reported in \citet{Veresvarska2024}, while the remaining targets are shown in Figure \ref{fig:LC}. WZ Sge shows a greater amplitude of variability akin to an envelope in sector 54. This is also seen in CP Pup \citep{Bruch2022,Veresvarska2024} and is thought to be an instrumental effect as it is displayed in multiple objects in the same sector, e.g. TIC 1909750039, TIC 1688054795 and TIC 1713691071 in sector 54 all show a similar envelope. This is not corrected for, but the analysis as described in Section \ref{s:Methods} has been conducted on sectors 41 and 54 separately. Since it has been found that the results are robust to any changes between the 2 sectors, it is assumed that the effect of the envelope on the results is negligible.

For targets with more than 1 sector (WZ Sge and T Pyx) the combination of the different sectors using un-calibrated \tess\ absolute photometry is possible because the combination happens in frequency domain, where separate power spectra are combined after they have been rms normalised. In such a case the relative variability in separate sectors is conserved and does not affect the overall result. As a test we have conducted the same analysis on un-calibrated \tess\ data of CP Pup and found no significant differences.

\begin{table*}
	\centering
	\caption{Summary of the \tess\ data of WZ Sge, CP Pup, GW Lib, T Pyx and V3101 Cyg with sector numbers and dates. The used exposure time of \tess\ sectors is noted for each sector, keeping it constant for each object. The optimal segment length $p_{0}$ used for constructing TPSs in \ref{ss:PSDs} is noted as well as whether the \tess\ data is converted to flux in mJy or kept in \tess\ default e$^{-}$s$^{-1}$.}
	\label{tab:obs}
	\begin{tabular}{lcccccr} 
		\hline
		Name & TIC & Sector & Dates & $t_{exp}$ (s) & $p_{0}$ (d) & Calibration \\
		\hline
		WZ Sge & 86408822 & 41 & 20/07/21 - 20/08/21 & 20 & 6 & $-$ \\
          &   & 54 & 09/07/22 - 05/08/22 & 20 & 6 & $-$ \\
        CP Pup & 14560527 & 7 & 08/01/19 - 01/02/19 & 120 & 10 & \asassn\ \\
         &  & 8 & 02/02/19 - 27/02/19 & 120 & 10 & \asassn\ \\
         &  & 34 & 14/01/21 - 08/02/21 & 120 & 10 & \asassn\ \\
         &  & 35 & 09/02/21 - 06/03/21 & 120 & 10 & \asassn\ \\
         &  & 61 & 18/01/23 - 12//02/23 & 120 & 10 & \asassn\ \\
        GW Lib & 225798235 & 38 & 29/04/21 - 26/05/21 & 20 & 10 & $-$ \\
        T Pyx & 17897279 & 35 & 09/02/21 - 06/03/21 & 120 & 7 & $-$ \\
          &  & 62 & 12/02/23 - 10/03/23 & 120 & 7 & $-$ \\
        V3101 Cyg & 1974089138 & 55 & 05/08/22 - 01/09/22 & 120 & 10 & $-$ \\
		\hline
	\end{tabular}
\end{table*}

\section{Methods and Analysis}
\label{s:Methods}

To draw an analogy to XRBs and the QPOs they exhibit, we attempt to reproduce their analysis as closely as possible. This involves construction of a Time-averaged Power Spectrum (TPS, as described in detail in Section \ref{ss:PSDs} and shown in Figure \ref{fig:PSD}) instead of inspecting a non-averaged and non-binned PSD (Figure \ref{fig:PSD_lin}). The details of the broad-band PSD fits are described in Section \ref{ss:PSDfit}, where the broad-band components of the TPS are fitted with zero-centred Lorentzians akin to XRBs.  Section \ref{ss:break} discusses the significance of the lowest frequency broad-band component of the fit from Section \ref{ss:PSDfit}.

\subsection{Broad-band Structure of the Time-averaged Power Spectrum}
\label{ss:PSDs}

We here characterise the broad-band variability and QPOs using TPS as is conventionally done when studying XRBs \citep{Belloni2002,Ingram2019}. TPS are usually constructed by separating the light curve into non-overlapping segments of equal length and computing the Fourier transform for each segment. The resulting power spectra are then averaged and re-binned onto a coarser frequency grid to reduce uncertainty. The error on the normalised power in each frequency bin corresponds to the standard error on the mean. In XRBs the nature of X-ray timing requires the use of Fast-Fourier transform (FFT). Here we adopt the Lomb-Scargle \citep{Lomb1976,Scargle1998} algorithm as the data obtained from \tess\ is not strictly evenly sampled due to gaps in the data from data downlinks and exclusion of some data points due to instrumental effects or other anomalous events. The Lomb-Scargle implementation used is \texttt{Astropy} v.5.3.4 with the limiting upper frequency set to the Nyquist frequency based on the sampling in Table \ref{tab:obs} and the lower frequency set to $3\times(L)^{-1}$, where $L$ is the length of the light curve. The oversampling factor is set to 1. The resulting power spectrum is normalised so that the total power of the power spectrum corresponds to $\frac{\sigma}{rms^{2}}$, where $\sigma$ is the variance of the light curve and $rms$ the root mean square. The resulting TPSs of WZ Sge, CP Pup, GW Lib, T Pyx and V3101 Cyg are shown in Figure \ref{fig:PSD}. 

In Figure \ref{fig:PSD} the segment length that is used to construct the TPSs is specified for each object in Table \ref{tab:obs}. After testing the robustness of the fit in Section \ref{ss:PSDfit} by varying the segment length, the length with the best visualisation of the PSD features were chosen. A similar method was used to select the number of frequency bins ($N_{bins} = 90$), ensuring  that the frequency resolution remains high enough to clearly distinguish QPOs from the broad-band PSD structure. This results in a rebinning factor of 270 for WZ Sge, 45 for T Pyx, 468 for GW Lib, 80 for V3101 Cyg and CP Pup.

The TPSs are also cleaned of the orbital signal, where present. This is not necessary for XRBs as most  XRBs have orbital signals on the order of several hours, where the signal does not interfere with QPOs present on timescales of $\sim$seconds. The power originating from the orbital motion is removed after computing the Lomb-Scargle for each segment. For each segment the peak of the orbital period is identified and $\sim$25 points on each side of the peak are removed, varying the number as necessary according to the strength and width of the signal. The same process is then repeated for 10 following harmonics of the orbit, 20 in the case of WZ Sge. The number of removed points was chosen after visual inspection to confirm that this would not remove any variability intrinsic to the QPO or significantly affect the broad-band PSD. This process is chosen over pre-whitening the light curve to avoid any possible changes to the underlying broad band-variability. Figure \ref{fig:PSD} displays the TPSs with the orbital variability already removed for WZ Sge, T Pyx, and V3101 Cyg. The linear PSD with the orbital periods included can be seen in Figure \ref{fig:PSD_lin}. The process was not necessary for GW Lib due to the low inclination of the system causing the orbital period to be only detectable spectroscopically \citep{Thorstensen2002}. Due to the lack of any periodic signals in the power spectrum of CP Pup and the uncertain nature of its orbital period \citep{Veresvarska2024}, no orbital period variations were removed.

\subsection{Time-averaged Power Spectra}
\label{ss:PSDfit}

\begin{figure*}
	\includegraphics[width=\textwidth]{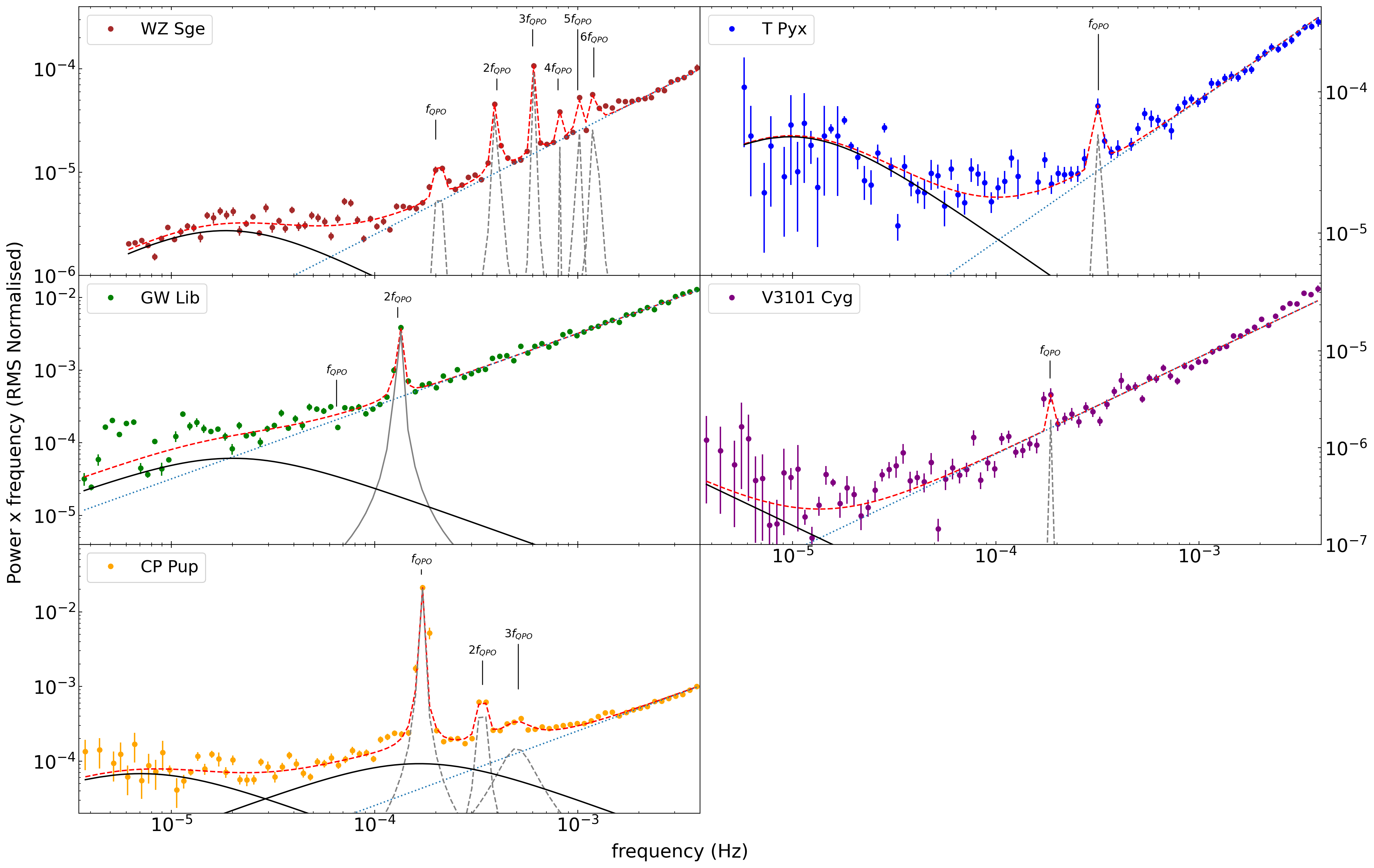}
    \caption{TPS of WZ Sge, GW Lib, T Pyx, V3101 Cyg and CP Pup averaged on the segment lengths specified in Table \ref{tab:obs} with 90 logarithmicaly spaced bins. The empirical fit of the PSD is included here and comprises a Poisson noise power law (dotted line) in the high frequencies, one (in the case of WZ Sge, GW Lib, T Pyx and V3101 Cyg) or two (CP Pup) zero-centred Lorentzians (solid line) representing the broad band PSD structure and Lorentzians representing the QPOs and their harmonics (dashed line).}
    \label{fig:PSD}
\end{figure*}

The broad-band structure of systems that accrete via an accretion disc can usually be fitted well with a sum of Lorentzian components \citep{Belloni2002}:

\begin{equation}
    P_{L} \left( \nu \right) = \frac{r^{2} \Delta}{\pi} \frac{1}{\Delta^{2} + \left( \nu - \nu_{0} \right)^{2}},
\label{eq:Lorentz}
\end{equation}

\noindent where $P_{L} \left( \nu \right)$ is the RMS normalised power, $\Delta$ the half width half maximum (HWHM) and $r$ a normalisation factor proportional to the integrated fractional rms. $\nu_{0}$ represents the centroid frequency, which is set to 0 in the case of zero-centred Lorentzians. In the case of non-zero centred Lorentzians, where $\nu_{0} > 0$, the frequency of the peak of the PSD feature is given by:

\begin{equation}
    \nu_{max} = \sqrt{\nu_{0}^{2} + \Delta^{2}}.
\label{eq:numax}
\end{equation}

\noindent Therefore the PSDs in Figure \ref{fig:PSD} are fitted with a number of non-zero centred Lorentzians corresponding to the number of quasi-periodic signals and their harmonics, plus one or two zero-centred Lorentzians representing the broad-band structure of the PSD. Two zero-centred Lorentzians were necessary only in the case of CP Pup. This is most likely due to the Poisson noise (see below) contributing mostly at high frequencies and intrinsic broad-band variability becoming dominant at $\sim 10^{-3}$ Hz. This can be seen in Figure \ref{fig:PSD} at the frequency where the Poisson noise component overtakes the zero-centred Lorentzian broad-band component. By contrast, this changeover occurs at $\sim 10^{-4}$ Hz for WZ Sge, GW Lib and V3101 Cyg.

The Poisson noise is another component of the PSD that needs to be considered. It can be represented by a constant, so that $P \left( \nu \right) = A$, where $A$ represents the amplitude of the noise. 

The PSD is fitted with all the free parameters for each object listed in Table \ref{tab:PSDfit}. The overall PSD model is hence expressed as:

\begin{equation}
    P_{\nu} = \sum_{i=0}^{N_{QPO}} P_{L} \left( r_{i}, \Delta_{i}, \nu_{0,i}, \nu \right) + \sum_{j=0}^{N_{L_{0}}} P_{L_{0}}\left(r_{j},\Delta_{j}, \nu \right) + A.
\label{eq:PSD}
\end{equation}

\noindent Here, $P_{L}$ is a non-zero centred Lorentzian representing the quasi-periodic signals and their harmonics, $P_{L_{0}}$ corresponds to the zero-centred Lorentzians and $A$ accounts for the Poisson noise. 

The best fit parameters with the corresponding errors are shown in Table \ref{tab:PSDfit}. The best fit is determined using the Levenberg-Marquardt least-squares fitting method with keeping all parameters free. Given the number of free parameters in the fit (e.g. 21 for WZ Sge) it is not feasible to conduct a full parameter sweep to determine the true confidence contours. Instead an assumption that the correlation between individual parameters is not dominant is made. In such a case, a rough estimate of errors can be made by fixing all parameters to the best fit value and varying a single parameter to determine the one dimensional confidence contours of 99.7$\%$. This technique was adapted from QPO fitting procedures in XRBs in cases where no strong correlations between PSD components can be assumed. This is applicable here  due to the narrowness of the QPOs and the low number of zero-centred Lorentzians required to fit the individual PSDs. Further complications may however arise from  the unevenly spaced data causing the frequency bins to not be strictly independent. Despite this it is important to note that the method cannot provide the exact 99.7$\%$ confidence contours, but can serve as a lower limit.

\begin{table*}
	\centering
	\caption{Results of the empirical fit to the PSDs from Figure \ref{fig:PSD}. The independent columns correspond to different objects and the rows correspond to model components: A $-$ Poisson noise amplitude, $r_{i}$ $-$ integrated fractional rms of a given Lorentzian, $\Delta_{i}$ $-$ half width half maximum of a given Lorentzian, $\nu_{0,i}$ $-$ centroid frequency of a given Lorentzian (only Lorentzians representing QPOs and their harmonics have this component). The Q value for each Lorentzian representing a QPO is given as well and is denoted by $Q_{i}$.} Broad-band component Lorentzians are zero-centred and their $\nu_{0}$ = 0 by default and is not denoted here. $^{*}$ denotes the lowest frequency zero-centred Lorentzian used in Figure \ref{fig:QPOsbreakXRBWD} and \ref{fig:QPOsXRBWD}.
	\label{tab:PSDfit}
	\begin{tabular}{lccccr} 
		\hline
		   & WZ Sge & CP Pup & GW Lib & T Pyx (s) & V3101 Cyg \\
		\hline
		A $(\times 10^{-2})$& 2.5$^{+0.2}_{-0.2}$ & 25$^{+2}_{-2}$ & 56$^{+5}_{-5}$ & 8.5$^{+0.8}_{-0.8}$ & 8.7$^{+0.8}_{-0.8} \times$ 10$^{-2}$ \\
        $r_{1}$ $(\times 10^{-2})$& 1.0$^{+0.3}_{-0.4}$ & 59$^{+10}_{-10}$ & 6$^{+4}_{-6}$ & 7$^{+3}_{-7}$ & 1.1$^{+0.8}_{-1.1} \times$ 10$^{-1}$ \\
        $\Delta_{1}$ $(\times 10^{-8})$& 4$^{+3}_{-3}$ & 70$^{+30}_{-30}$ & 2$^{+4}_{-2}$ & 0.7$^{+0.9}_{-0.7}$ & 0.1$^{+0.2}_{-0.1}$\\
        $\nu_{0,1}$ $(\times 10^{-4})$& 2.07$^{+0.05}_{-0.04}$ & 1.70$^{+0.1}_{-0.05}$ & 1.33$^{+0.01}_{-0.05}$ & 3$^{+2}_{-2}$ & 1.8665$^{+1}_{-0.0009}$\\
        $Q_{1}$ & $\sim$ 2600 & $\sim$ 120 & $\sim$ 3300 & $\sim$ 21000 & $\sim$ 93000 \\
        $r_{2}$ $(\times 10^{-2})$& 0.21$^{+0.05}_{-0.06}$ & 10$^{+2}_{-2}$ & 1.0$^{+0.3}_{-0.4}$ & 1.7$^{+0.2}_{-0.2}$ & 1.9$^{+0.8}_{-1.5}$\\
        $\Delta_{2}$ $(\times 10^{-5})$& 0.7$^{+0.7}_{-0.4}$ & 7$^{+4}_{-4} \times$ 10$^{-3}$ & 1.6$^{+2.8}_{-1.2}$ $^{*}$ & 1.0$^{+0.3}_{-0.5}$ $^{*}$ & 0.13$^{+0.13}_{-0.13}$ $^{*}$\\
        $\nu_{0,2}$ $(\times 10^{-4})$& 4.0$^{+0.1}_{-0.1}$ & 3.4$^{+0.1}_{-0.1}$ & $-$ & $-$ & $-$\\
        $Q_{2}$ & $\sim$ 30 & $\sim$ 2500 & $-$ & $-$ & $-$ \\
        $r_{3}$ $(\times 10^{-2})$& 1.1$^{+0.3}_{-0.4}$ & 1.0$^{+0.2}_{-0.3}$ & $-$ & $-$ & $-$\\
        $\Delta_{3}$ $(\times 10^{-4})$& 1.4$^{+0.7}_{-0.7} \times$ 10$^{-3}$ & 1.0$^{+2}_{-0.5}$ & $-$ & $-$ & $-$\\
        $\nu_{0,3}$  $(\times 10^{-4})$& 6.1$^{+0.2}_{-0.2}$ & 4.9$^{+0.6}_{-0.7}$ & $-$ & $-$ & $-$\\
        $Q_{3}$ & $\sim$ 2200 & $\sim$ 2.5 & $-$ & $-$ & $-$ \\
        $r_{4}$ $(\times 10^{-2})$& 13$^{4}_{-6}$ & 2.4$^{+0.2}_{-0.3}$ & $-$ & $-$ & $-$\\
        $\Delta_{4}$ $(\times 10^{-4})$& 1.5$^{+1}_{-1} \times$ 10$^{-5}$ & 1.7$^{+8}_{-5}$ & $-$ & $-$ & $-$\\
        $\nu_{0,4}$  $(\times 10^{-4})$& 8$^{+4}_{-7}$ & $-$ & $-$ & $-$ & $-$\\
        $Q_{4}$ & $\sim$ 2.7 & $-$ & $-$ & $-$ & $-$ \\
        $r_{5}$ $(\times 10^{-2})$& 1.7$^{+0.5}_{-0.8}$ & 2.1$^{+0.4}_{-0.4}$ & $-$ & $-$ & $-$\\
        $\Delta_{5}$ $(\times 10^{-7})$& 1.0$^{+0.7}_{-0.7}$ & 70$^{+50}_{-30}$ $^{*}$ & $-$ & $-$ & $-$\\
        $\nu_{0,5}$  $(\times 10^{-4})$& 1.0$^{+0.2}_{-0.2}$ & $-$ & $-$ & $-$ & $-$\\
        $Q_{5}$ & $\sim$ 500 & $-$ & $-$ & $-$ & $-$ \\
        $r_{6}$  $(\times 10^{-2})$& 1.6$^{+0.5}_{-0.7}$ & $-$ & $-$ & $-$ & $-$\\
        $\Delta_{6}$  $(\times 10^{-7})$& 3$^{+2}_{-2}$ & $-$ & $-$ & $-$ & $-$\\
        $\nu_{0,6}$  $(\times 10^{-3})$& 1.22$^{+0.03}_{-0.02}$ & $-$ & $-$ & $-$ & $-$\\
        $Q_{6}$ & $\sim$ 200 & $-$ & $-$ & $-$ & $-$ \\
        $r_{7}$  $(\times 10^{-3})$& 4.1$^{+0.3}_{-0.3}$ & $-$ & $-$ & $-$ & $-$\\
        $\Delta_{7}$  $(\times 10^{-5})$& 1.9$^{+0.4}_{-0.3}$ $^{*}$ & $-$ & $-$ & $-$ & $-$\\
        \hline
        $\chi^{2}_{\nu}$ & 6.16 & 3.41 & 1.95 & 1.52 & 5.26 \\
        \hline
	\end{tabular}
\end{table*}

\subsection{Break Significance}
\label{ss:break}

The uncertainty on the low frequency broad-band feature (break) measured in Table \ref{tab:PSDfit} is difficult to constrain due to several factors. One relates to the logarithmic spacing of the frequency bins resulting in a decrease of the number of points in the lower frequency bins, increasing their error. Therefore it is necessary to verify the presence of any break as opposed to a red-noise related power-law.

To generate light curves that follow a specific power-law pattern, we use an algorithm similar to the one described by \citet{Veresvarska2023}, based on the method developed by \citet{Timmer1995}. Given a PSD model, the algorithm will generate a simulated PSD by scattering the amplitudes and phases around the PSD model. The randomized PSD is then transformed back into the time domain using an inverse Fourier transform, creating a light curve with the desired PSD shape.

In this case, the PSD shape is almost the same as the one outlined in Table \ref{tab:PSDfit}. However, the lowest frequency zero-centred Lorentzian is replaced by a power-law fit, as shown in Figure \ref{tab:PSDTKfit}, where the power-law is defined by $P_{power-law} = A \nu^{\alpha}$. The Lorentzians corresponding to the QPOs are removed, and the other parameters from Table \ref{tab:PSDfit} are kept constant. The values for $A$ and $\alpha$ are found using the Levenberg-Marquardt least-square method. The reduced $\chi^{2}_{\nu}$ for the fits are also noted in Table \ref{tab:PSDTKfit}. It is worth noting that all objects apart from V3101 Cyg have a significantly higher $\chi^{2}_{\nu}$ for the power-law fit than when the zero-centred Lorentzians are used.

The parameters from Table \ref{tab:PSDfit}, combined with the best-fit values of $A$ and $\alpha$, are used to generate artificial light curves. The components involved in this process are listed in Table \ref{tab:PSDTKfit}. In this context, $r_{last}$ and $\Delta_{last}$ refer to the parameters of the lowest frequency zero-centred Lorentzian for a given object in Table \ref{tab:PSDfit}. The errors shown in Table \ref{tab:PSDTKfit} are calculated in the same way as those in Table \ref{tab:PSDfit}, as detailed in Section \ref{ss:PSDfit}.

\begin{table*}
	\centering
	\caption{Results of the empirical fit to the PSDs from Figure \ref{fig:TK}. The independent columns correspond to different objects and the lines correspond to model components: A $-$ power-law amplitude, $\alpha$ $-$ power-law index.}
	\label{tab:PSDTKfit}
	\begin{tabular}{lccccr} 
		\hline
		   & WZ Sge & CP Pup & GW Lib & T Pyx (s) & V3101 Cyg \\
		\hline
        $A$ & 9$^{+2}_{-2}\times$ 10$^{-10}$ & 2$^{+8}_{-2}\times10^{-8}$ & 1$\times 10^{-4}$$^{+4}_{-1\times 10^{-4}}$ & 5$^{+2}_{-2}$$\times$ 10$^{-9}$ &1$ ^{+13}_{-1}\times 10^{-13}$ \\
        $\alpha$ & -1.756$^{+0.003}_{-0.004}$ & -1.684$^{+0.005}_{-0.004}$ & -0.883$^{+0.004}_{-0.003}$ & -1.778$^{+0.004}_{-0.005}$ & -2.248$^{+0.010}_{-0.009}$\\
        \hline
        $\chi^{2}_{\nu}$ & 163 & 11.6 & 14.5 & 5.2 & 5.00 \\
        \hline
	\end{tabular}
\end{table*}

\begin{figure*}
	\includegraphics[width=\textwidth]{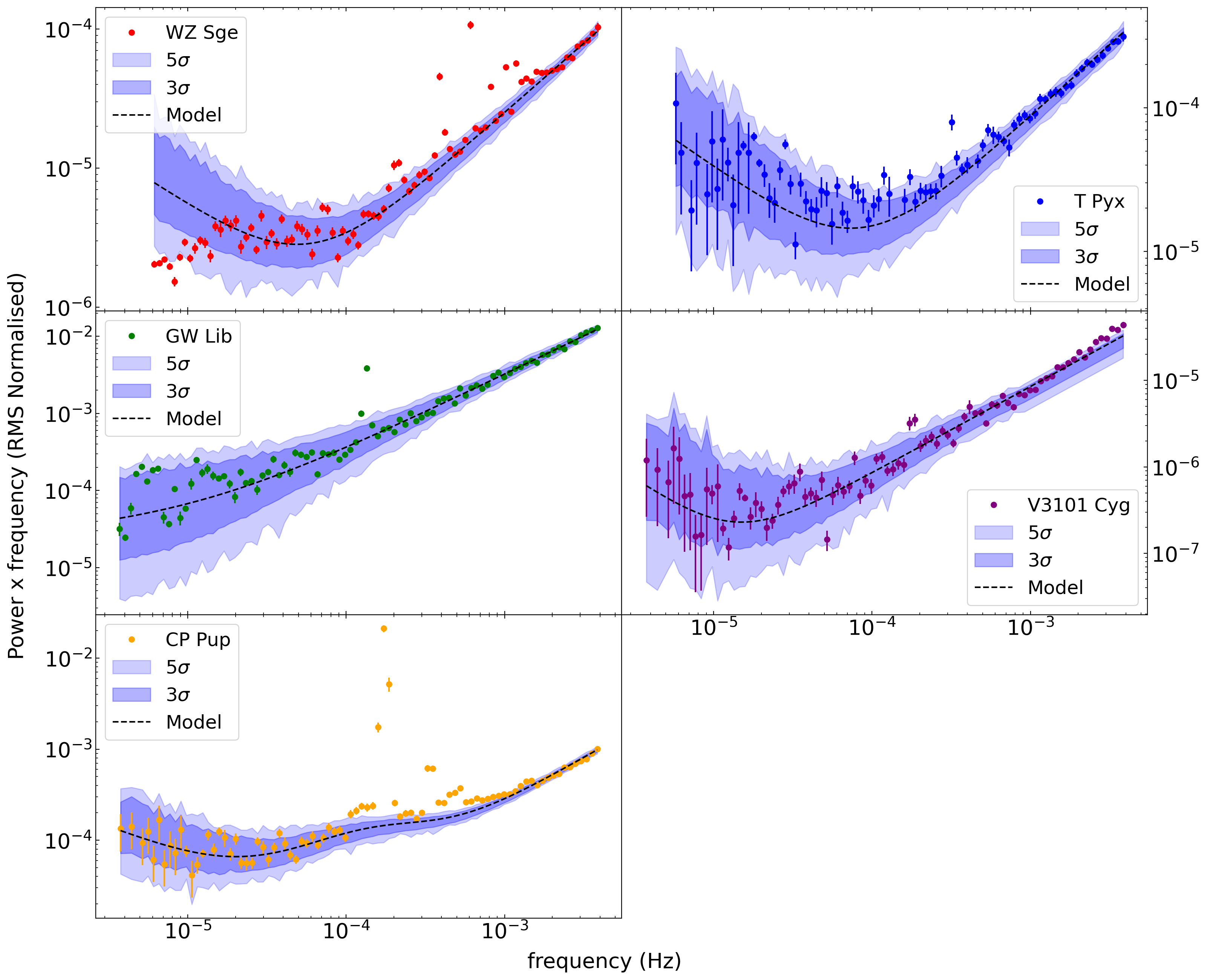}
    \caption{PSDs as shown in Figure \ref{fig:PSD}. Shaded regions show confidence contours from \citet{Timmer1995} algorithm. Dark shaded region corresponds to 3$\sigma$ significance and light shaded region to 5$\sigma$ significance. The data points outside of the shaded regions boundaries denote that the input power law (dashed line) cannot explain the underlying PSD shape to 3$\sigma$ or 5$\sigma$ significance.The QPOs are outside of the 3$\sigma$ and 5$\sigma$ significance since their respective Lorentzians have not been included in the model. This showcases their $>>5\sigma$ significance above the underlying power-law.}
    \label{fig:TK}
\end{figure*}

The process of generating artificial light curves is repeated 1000 times, resampling the final output on the time stamps of the original \tess\ light curves, reproducing identical PSDs as in Figure \ref{fig:PSD}. All the points in each frequency bin are examined and fitted with a $\chi^{2}$ distribution. The optimal fit of the $\chi^{2}$ distribution is obtained again with the Levenberg-Marquardt least-square method. The 3$\sigma$ and 5$\sigma$ confidence contours corresponding to 93.32$\%$ and 99.977$\%$ levels are extracted and shown in Figure \ref{fig:TK}. The envelopes determine the significance level of RMS in each frequency bin independently. Therefore any observed trend in deviation from a certainty envelope signifies a higher level of confidence that the input power law does not represent a viable model for the PSD.

The resulting certainty envelopes are shown for all objects in Figure~\ref{fig:TK}. It is clear that WZ Sge contains a feature, that cannot be explained by a simple power-law, with a certainty over 5$\sigma$. Other deviations from the certainty envelope such as the QPO with its harmonics shows the significance of the QPO signal. Other such features similar to the bump at 7$-$8 $\times$ 10$^{-5}$ Hz represent broad-band features that were not included in the PSD fit. Similar deviations at frequencies $>$ 10$^{-3}$ Hz are influenced by the remaining harmonics of the QPO that were not fitted as their corresponding frequency bins did not allow for separation of individual harmonics.

No significant deviation from the 5$\sigma$ significance envelope is seen in any other objects apart from WZ Sge. The long-term TPS of AWDs are known to have a complex structure requiring multiple Lorentzians \citep{Scaringi2012a}. However, using the 3$\sigma$ and 5$\sigma$ confidence contours it is difficult to say to what extent this structure translates to the objects here, apart from the clear example of WZ Sge. Some structure is hinted at in GW Lib, but it is obstructed by the dominant Poisson noise component and limited amount of data available (compared to WZ Sge and MV Lyr in \citealt{Scaringi2012a}).  

Despite not being able to probe low enough frequencies to test the presence of a low frequency break with $\gtrsim 5 \sigma$ significance in all the systems, the obtained significance combined with the $\chi^{2}$ comparison of the 2 fits, shows evidence of self-similar empirical features to those in broad-band PSDs in XRBs. A zero-centred Lorentzian is used across all 5 systems, since the majority of the targets show a significantly lower $\chi^{2}$ for the zero-centred Lorentzian fit. However, to err on the cautious side, the uncertainties on the break component are treated as an upper limit in all targets apart from WZ Sge.

\section{Results}
\label{s:results}

In this section the main results are outlined. The new QPOs in AWDs are reported in terms of the existing QPOs and broad-band low frequency break correlation from \citet{Wijnands1999}. The link to QPOs in XRBs is drawn from the broadness of the signal in the power spectrum as seen in Figure \ref{fig:PSD_lin} and explained further in Section \ref{ss:QPOs_ints}. Another important characteristic of the new QPOs in AWDs is that it is the first instance in which QPOs in AWDs are reported to show harmonics as seen in Figure \ref{fig:PSD}.

As mentioned before in Section \ref{s:Intro}, type-C and HBO QPOs in XRBs show a linear correlation with a low frequency break in their PSDs. The break corresponds to the nearest low zero-centred Lorentzian that is used in the fit of the broad-band PSD and was first reported in \citet{Wijnands1999} with an overview available in \citet{Ingram2019}.

\begin{figure}
	\includegraphics[width=\columnwidth]{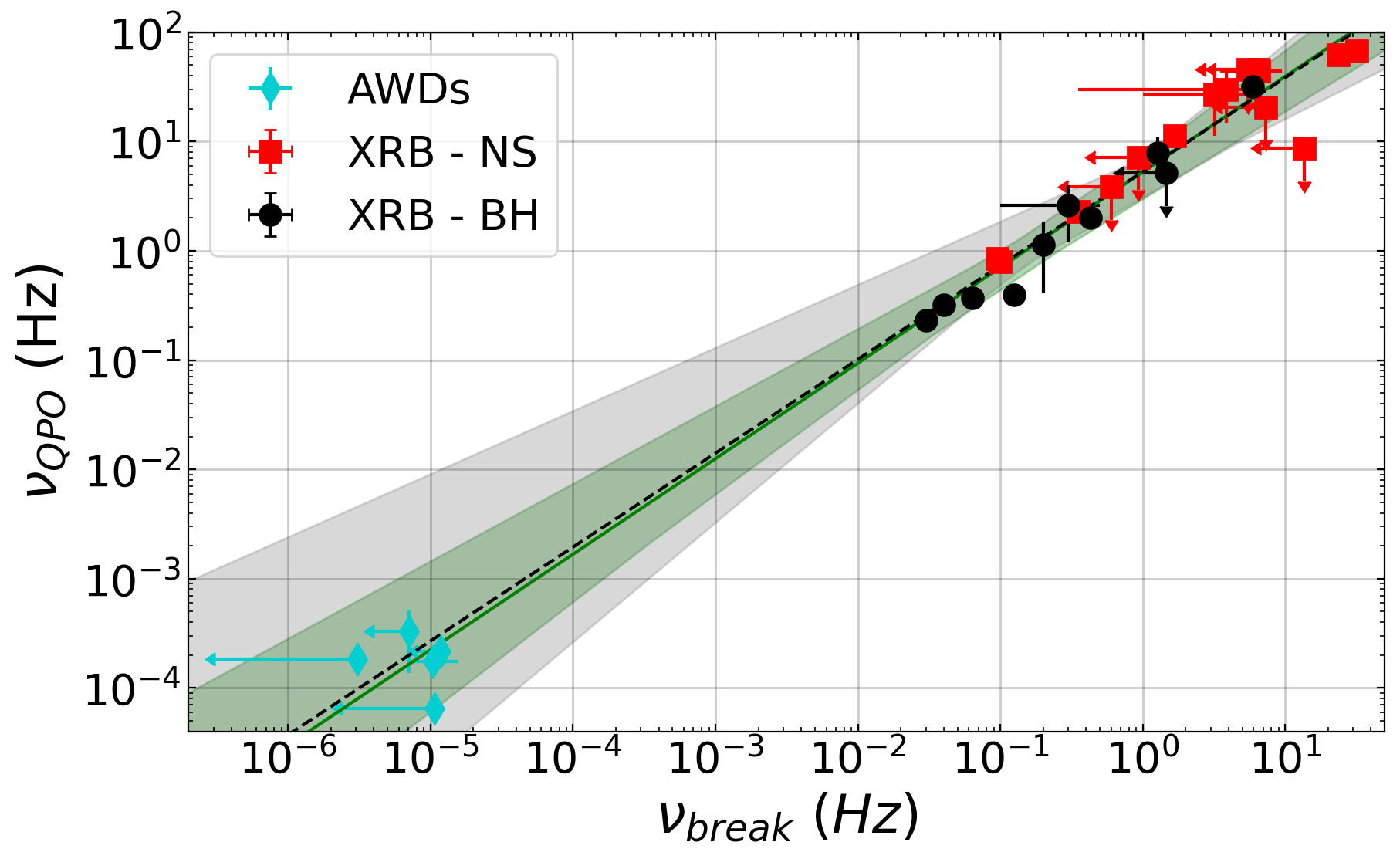}
    \caption{QPO frequency as a function of the break frequency. The observed QPOs with their corresponding low frequency breaks are denoted for AWDs in diamonds, for NS XRBs in squares and in filled circles for BH XRBs. For AWDs these are taken from the empirical fit in Section \ref{ss:PSDfit} and Table \ref{tab:PSDfit}. The QPO frequency is the peak frequency as denoted in Equation \ref{eq:numax} using $\Delta_{1}$ and $\nu_{0,1}$. The only exception is GW Lib where the QPO frequency is taken as half that, since only the 1$^{st}$ harmonic of the QPO is detected in \tess\ with its true fundamental being reported in \citet{Chote2021MNRAS.502..581C}. The break frequencies are denoted in Table \ref{tab:PSDfit} by $^{*}$ for each object.The solid line represents the fit to the data from linear least-squares regression and the shaded darker area represents the uncertainty on the fit. The dashed line represents the fit to the data when AWDs are excluded, with the uncertainty being represented by the lighter shaded region.}
    \label{fig:QPOsbreakXRBWD}
\end{figure}

The results of taking the lowest zero-centred Lorentzian component from Table \ref{tab:PSDfit} (denoted by $^{*}$) and plotting it against the characteristic QPO frequency as defined in Equation \ref{eq:numax} is shown in Figure \ref{fig:QPOsbreakXRBWD}. Since for GW Lib the fundamental of the QPO signal reported by \citet{Chote2021MNRAS.502..581C} is not detected in \tess\ , half of the 2$^{nd}$ harmonic frequency ($\Delta_{1}$ and $\nu_{0,1}$ from Table \ref{tab:PSDfit}) is used instead. Following from Section \ref{ss:break} only upper limits are used for the zero-centred Lorentzian in all systems apart from WZ Sge and are denoted by arrows. BH and NS QPOs with their corresponding low-frequency broad-band components (breaks) from Table 1 in \citet{Wijnands1999} are also shown.

\begin{figure}
	\includegraphics[width=\columnwidth]{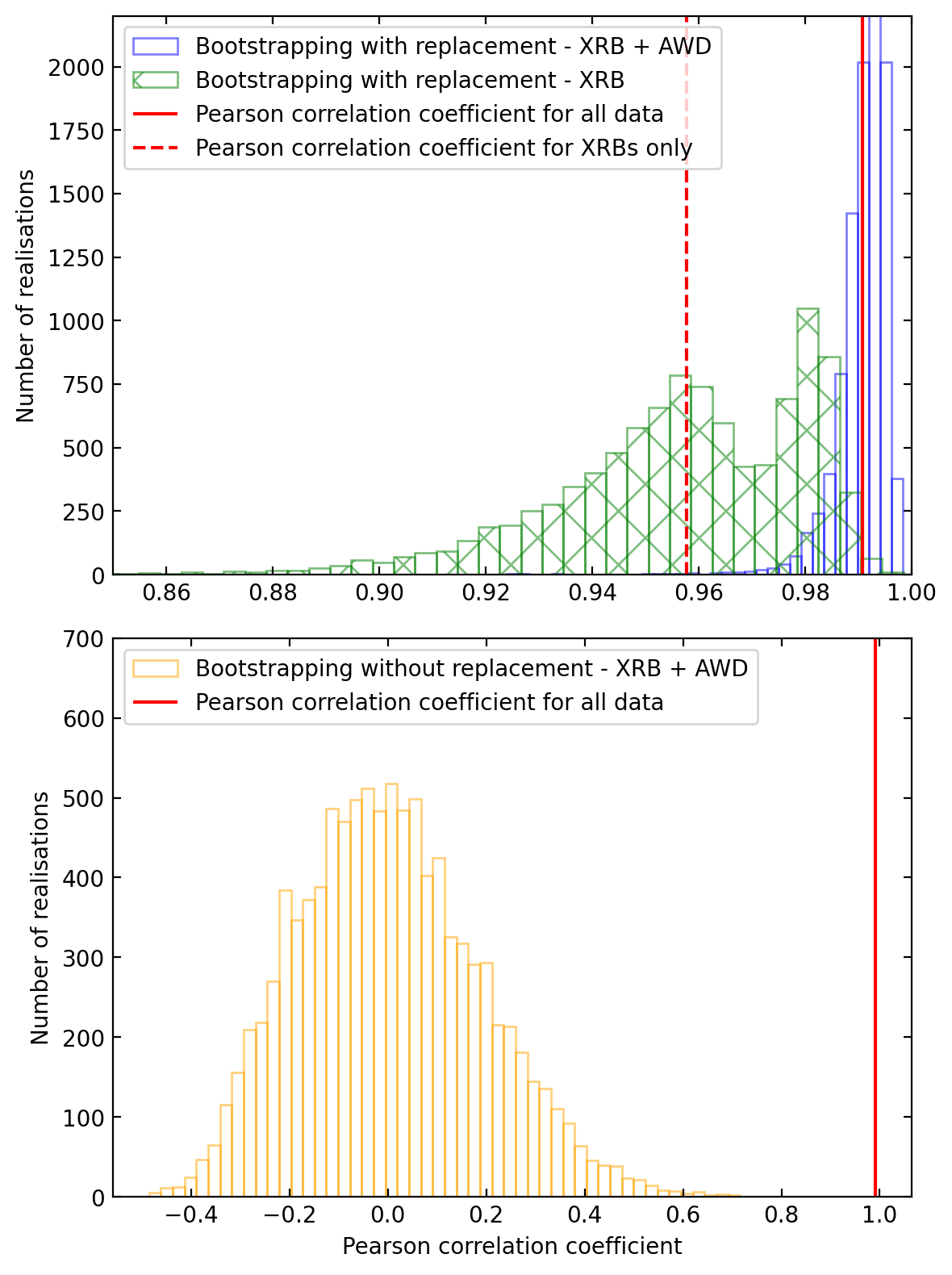}
    \caption{\textit{Top}: Distribution of the Pearson correlation coefficient for the bootstrapping of data from Figure \ref{fig:QPOsbreakXRBWD} with replacement. The clear distribution corresponds to bootstrapping of all data, whilst the hatched distribution only drew from the XRB sample from Figure \ref{fig:QPOsbreakXRBWD}. The solid line represents the Pearson correlation coefficient of of all data in Figure \ref{fig:QPOsbreakXRBWD} and the dashed line corresponds to the value of the coefficient when AWDs are excluded.\textit{Bottom}: The distribution of the Pearson correlation coefficient resulting from bootstrapping the data from Figure \ref{fig:QPOsbreakXRBWD} without replacement and by randomising one of the axes. The zero-centred distribution shows the 99.99$\%$ significance of the correlation from Figure \ref{fig:QPOsbreakXRBWD}, with the solid line marking the value of the Pearson correlation coefficient of the data.}
    \label{fig:bootstrap}
\end{figure}

The points representing AWDs are clearly following the same empirical relation as the XRBs from \citet{Wijnands1999}. The AWDs are several orders of magnitude lower in both QPO and break frequency. To demonstrate the significance of this correlation a linear fit is obtained using linear least-squares regression as implemented in \texttt{Scipy}. A Pearson correlation coefficient of 0.991 is obtained for all data shown in Figure \ref{fig:QPOsbreakXRBWD} and 0.805 for XRBs only. The resulting linear trend is shown in Figure \ref{fig:QPOsbreakXRBWD} as a solid line for all data and as a dashed line for solely XRBs. Bootstrapping is used to test the robustness of this correlation. This consists of randomising the QPO frequencies for corresponding break frequency values. In doing so, no correlation is expected and the Pearson correlation coefficient is expected to be $\sim$ 0. Repeating this process $10^{4}$ times then shows that the correlation is significant to 99.99\% confidence. The resulting distribution of the Pearson correlation coefficient is then shown in the bottom panel of Figure \ref{fig:bootstrap}. To constrain the uncertainty on these fits, N data points are randomly selected with replacement from the data set in Figure \ref{fig:QPOsbreakXRBWD}. The linear regression fit is then repeated and the Pearson coefficient is computed. The process is repeated $10^{4}$ times resulting in the mean Pearson coefficient of 0.991$\pm$0.005, a value consistent with the original Pearson correlation coefficient. This process is repeated for all data points in Figure \ref{fig:QPOsbreakXRBWD} as well as only XRB points with the resulting distributions shown in Figure \ref{fig:bootstrap}. The uncertainty in the linear regression fits performed during bootstrapping is shown in Figure \ref{fig:QPOsbreakXRBWD} as the lightly shaded regions for all data points and for XRBs only. The correlation can hence be expressed as $log(\nu_{QPO}) = 0.87 \pm 0.03 log(\nu_{break}) + 0.72 \pm 0.05$ for all data points and $log(\nu_{QPO}) = 0.86 \pm 0.06 log(\nu_{break}) + 0.73 \pm 0.05$ for XRBs only.

\section{Discussion}
\label{s:Discussion}

In this Section the implications of the QPO model as a potential tool for spin period and magnetic field strength measurement for weakly magnetised AWDs is discussed in Section \ref{ss:QPOmodel_disc}. In Section \ref{ss:QPO_break_disc} the QPO and break correlation for XRBs is discussed in the framework of AWDs and the implications for the driving mechanisms of the QPOs.

\subsection{Possible interpretations of QPOs in AWDs}
\label{ss:QPOs_ints}

There are numerous signals which are present in AWDs in the frequency range displaying the QPO signals shown in Figure \ref{fig:PSD}. 

The PSDs in Figure \ref{fig:PSD} are already cleaned of any variability related to the orbital period of the systems. Inspecting the nature of the signals in the  non-averaged and non-binned power spectra in Figure \ref{fig:PSD_lin}\, we find that the signals are phenomenologically different from the coherent signals usually present in AWDs, such as the orbital period or spin. The main difference being the power peaks of the signals are broader than coherent periods. Where present, the signal from the orbital period is also indicated in Figure \ref{fig:PSD_lin}.

Another type of coherent signal present in AWDs is usually due to negative and positive superhumps and their associated fundamental signals. In that case, the superhumps are present relatively close to the orbital period of the system. For a positive superhump, $\nu_{SH}^{+}$ < $\nu_{orb}$, and the signal is associated with the tidal stresses exerted by the secondary on the disc. These cause the disc to become eccentric and undergo apsidal precession in the prograde direction \citep{Lubow1991}. In this scenario, only GW Lib and V3101 Cyg would qualify. In neither system is a fundamental frequency of the signal found at low frequencies. For the case of negative superhumps, where $\nu_{SH}^{-}$ > $\nu_{orb}$, the signal is thought to be caused by the retrograde nodal precession of a tilted accretion disc \citep{Wood2009}, resulting in a 3:1 resonance. This is viable for WZ Sge and T Pyx. However, in the case of WZ Sge, CP Pup, GW Lib, and T Pyx, the QPO signal shows distinct harmonics, making a superhump explanation unfeasable. The presence of harmonics also rules out any other possibility of the signals being related to a beat between the orbit and spin. 

Another peculiar feature of the signals is that they are not entirely coherent, as shown in Figure \ref{fig:PSD_lin}. As opposed to the orbital signals, the QPO signals are broad, reaching a width of $\sim2 \times 10^{-5}$ Hz. They also exhibit slight variations in central frequency, amplitude and shape. This has already been reported for CP Pup \citep{Bruch2022,Veresvarska2024}. In GW Lib, the signal is revealed to be the first harmonic of the fundamental signal at $\sim$4 hours, interpreted as a quiescent superhump \citep{Chote2021MNRAS.502..581C} similar to the transitional feature in EQ Lyn \citep{Mukadam2013AJ....146...54M} and in V455 And \citep{Araujo-Betancor2005A&A...430..629A}. Since in \tess\ only the harmonic is detected, the fundamental is here inferred. This is similar behaviour to that of WZ Sge, where the subsequent harmonics are much stronger than the fundamental signal as is visible in Figure \ref{fig:PSD}. A parallel may be drawn with the sub-harmonics of QPOs observed in XRBs, where the fundamental signal is the strongest, not the lowest frequency \citep{Casella2005ApJ...629..403C}. The overall unusual behaviour of these signals suggests the necessity of an alternative explanation. 

QPOs in XRBs display strikingly similar properties. They show harmonics and time variability as well as general broadness of the signal itself \citep{Ingram2019}. This is apparent in the quality factor of the QPOs, which -- depending on the type of XRB QPO -- can be $Q \lesssim$ 3 or $Q \gtrsim$ 6. Due to the value of Q depending also on the variability of the signal during the observations, it may be expected that Q will be much larger for AWDs due to the slower variability time scales. This is indeed the case as shown by the reported Q values in Table \ref{tab:PSDfit}, where some QPOs have Q $\gtrsim$ 10000. This, whereas unusual for XRBs, would seem necessary for a signal to be detected above the Poisson noise dominated systems such as the ones shown here. However, despite such a high Q value the signals bear little similarity to other coherent signals such as the orbital period as described above and shown in Figure \ref{fig:PSD_lin}.

\subsection{Magnetically Driven Precession as QPO-driving Mechanism}
\label{ss:QPOmodel_disc}

QPOs in XRBs are generally explained by the relativistic effect of frame dragging modelled most commonly by Lense-Thirring precession \citep{Stella1998ApJ...492L..59S,Stella1999ApJ...524L..63S,Psaltis2000astro.ph..1391P,Fragile2001ApJ...553..955F,Ingram2009MNRAS.397L.101I}. While several possibilities have been presented, in particular for NS XRBs, Lense-Thirring precession is currently the most popular interpretation. However, since AWDs are indisputably non-relativistic, a different model is required to explain their QPOs. A model explored by \citet{Warner2002} proposes that a magnetospherically truncated accretion disc may give rise to QPOs from the interaction between twisted magnetic field lines and plasma in the disc. This may produce a dense and elevated region referred to as a "blob". The release of energy from the sudden reconnection of strained field lines within the blob is speculated to generate disturbances that propagate through the disc, potentially resonating with its natural oscillations and generating QPOs. In this scenario the frequency of the QPOs are associated with the recconection timescale estimated in \citet{Warner2002} to be on the order of $\sim$15 minutes for a non-resonant signal. The authors do not consider a resonant case as this would be too close to the orbital period and difficult to detect photometrically. The resulting frequency range thus corresponds to the QPOs and DNOs found by \citet{Warner2003MNRAS.344.1193W}, not the ones reported in this work which occur on longer timescales.

\begin{figure}
	\includegraphics[width=\columnwidth]{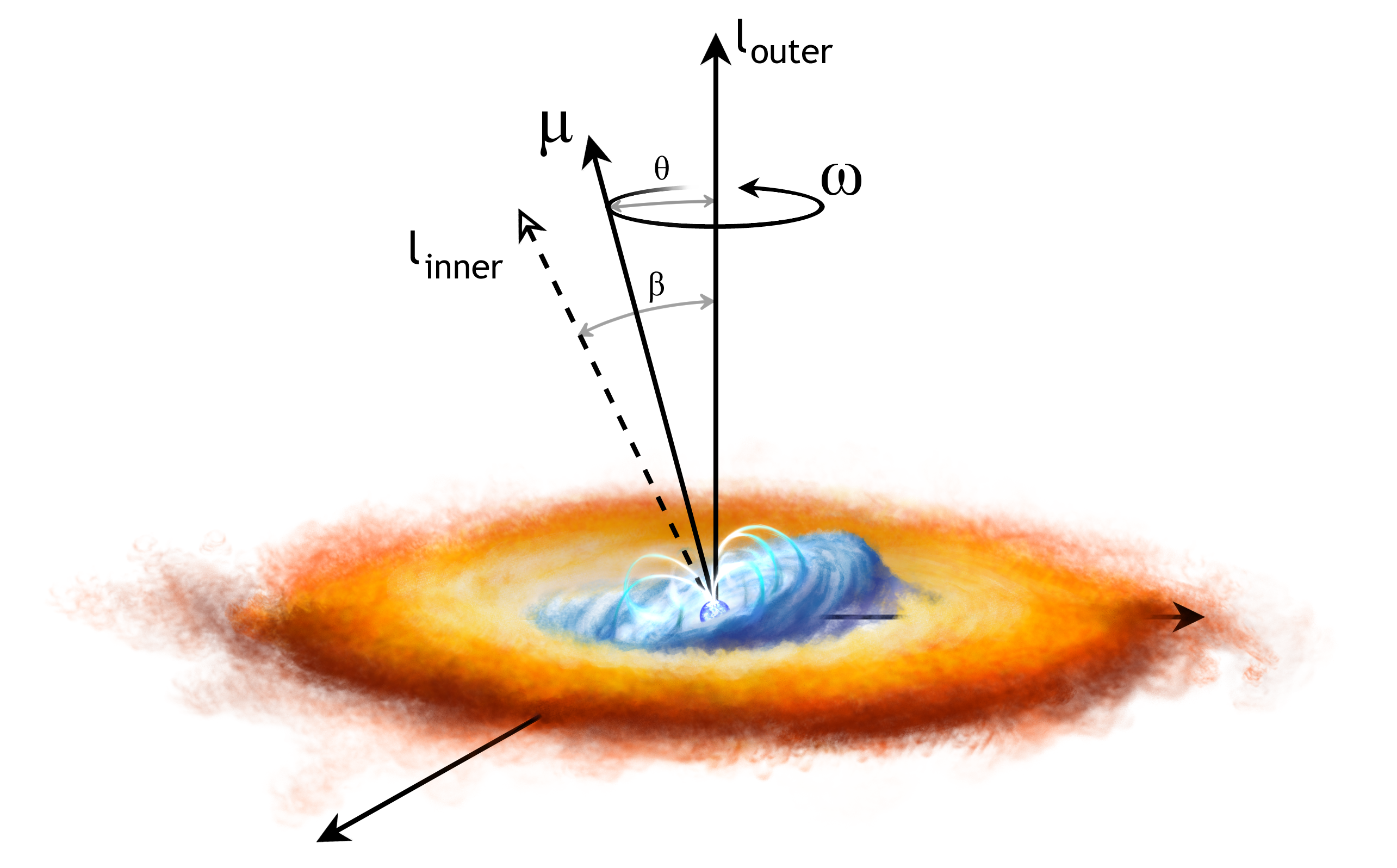}
    \caption{Diagram showing the disc geometry producing a QPO as described by \citep{Lai1999, Pfeiffer2004}. The accretor magnetic field vector $\mu$ and spin axis $\omega$ are misaligned to the angular momentum of the outer standard accretion disc ($l_{outer}$) leading to the precession of an inner part of the disc that is warped as a consequence ($l_{inner}$).}
    \label{fig:diagram}
\end{figure}

One alternative and intrinsically non-relativistic model for QPOs in neutron stars was proposed by \citet{Aly1990A&A...227..473A}, where the QPOs are driven by interactions between the disc and accretor magnetic field. However, under such framework the QPOs can only occur at a radius where the magnetic fields strengths of the accretor and disc are equal (Equation 2 in \citealt{Aly1990A&A...227..473A}), which may not be a feasible scenario for accreting white dwarfs. 

A further alternative model developed to explain QPOs in neutron stars and T Tauri stars was developed by \citet{Lai1999}. They suggest that QPOs may be associated with magnetically driven precession, a phenomenon where the orientation of the accretion disc surrounding a rotating magnetised star undergoes a periodic wobbling motion around the star's spin axis. This interaction between the disc and the star's rotating magnetic field induces a warping effect on the disc, causing it to deviate from its original equatorial configuration and to precess around the star. The precession torque arises from the interaction between the surface current on the disc and the horizontal magnetic field produced by the star's dipole. The nonlinear evolution of the disc's state is then further explored by \citet{Pfeiffer2004}, and the model's applications to NS low-frequency QPOs are discussed in \citet{Shirakawa2002a,Shirakawa2002b}.

A schematic sketch of the model geometry is shown in Figure \ref{fig:diagram}. In the framework of this model, as presented by \citet{Lai1999}, the star's rotational axis $\omega$ is tilted with respect to the disc's angular momentum (\textit{l}) by an angle $\beta$.

The QPO is generated by the precession of the inner disk around the spin axis of the accretor. \citet{Lai1999} approximate the precession frequency of the entire inner disk by multiplying the magnetic precession frequency of a ring at any specific characteristic radius $r$, $\nu_p(r)$, by a dimensionless constant $A$, such that the resulting QPO frequency can be expressed as

\begin{equation}
\nu_{QPO} = A \nu_p(r) = \frac{ A }{ 2\pi^3 } \frac{ \mu^2 }{ r^7~\Omega(r)~\Sigma(r) } \frac{ F(\theta) }{ D(r) },
\label{eq:nuQPO}
\end{equation}

\noindent where $\mu$ is the stellar magnetic dipole moment ($\mu=B R^3$, where $B$ is the field strength and $R$ is the stellar radius of the accretor), $\Omega$ is the Keplerian angular frequency, $\Sigma$ is the disk surface density, and $F(\theta)$ and $D(r)$ are dimensionless functions defined in Appendix C (where we discuss the model in more detail). Because of several degenerate parameters in this model, we here assume that the characteristic radius $r$ is the disk inner radius $r_{\rm in}$ which we set equal to both the magnetospheric radius and the coronation radius such that $r = r_{\rm in} = r_{\rm m} = r_{\rm co}$. A more detailed description of the model and its parameters is given in the Appendix \ref{a:model}. 

\subsubsection{Magnetic field strength, accretion rate, and spin estimation from QPOs in AWDs}
\label{sss:model_res}

Equation \ref{eq:nuQPO} depends on several free parameters. Amongst those are the accretor mass and size, the strength of the intrinsic magnetic field strength of the accretor $B$, the accretion rate of the system $\dot{M}$ and the dimensionless viscosity of the disc $\alpha$. A different combination of these parameters can yield different QPO frequencies which may occur at different radii. The radius of the QPO is given by the relation between the magnetospheric radius $r_{M}$ and the corotation radius $r_{CO}$. We here assume that $r_{in} = r_{M} = r_{CO}$, which allows to break some of the degeneracies in the model when inferring the accretor magnetic field and corresponding spin of the accretor. 

To explore the validity of the model we employ a set of fiducial parameters as several of these would otherwise remain unconstrained. The effect of varying the $B$ field and accretion rate is non-negligible, and no reasonable estimate can be made that would be applicable for all AWDs or NS XRBs. Therefore, we instead explore a range of these parameters for AWDs and NS XRBs as noted in Table \ref{tab:model_fid}. Hence, for a given magnetic field and accretion rate a specific combination of QPO frequency and spin period is obtained allowing for comparison to data for objects with QPO and spin period measurements.

Specifically for the purpose of comparing the model to the 2 observed QPOs in AWDs with spin period measurements, 2 separate values of viscosity are chosen. These correspond to the quiescent values of WZ Sge with $\alpha \sim$ 0.006 and $\alpha \sim$ 0.003  \citep{Howell1995ApJ...439..337H} for GW Lib, which are obtained from modelling of the recurrence timescales of the dwarf novae outbursts. For other objects the viscosity of WZ Sge as well as other values of fiducial parameters are used. The different fiducial parameters are due to the intrinsic differences between the systems. Despite both being dwarf novae with similar characteristics and recurrence times of the outbursts ($\sim$ years), the estimates of their viscosities from \citet{Howell1995ApJ...439..337H} provide an important constraint on otherwise fully unconstrained fiducial parameters. Ideally, to truly apply the magnetically driven QPO model to its full extent a precise measurement of all parameters is required. With this not feasible, it is necessary to use these viscosity estimates. Since the estimated accretion rate of GW Lib from the literature \citep{Hilton2007AJ....134.1503H} is much smaller than that of WZ Sge, the fiducial parameters of GW Lib are used in the treatment of the other objects showing QPO.

For the fiducial values of these parameters and each combination of $B$ and $\dot{M}$ there is a characteristic QPO frequency $\nu_{QPO}$ with an associated spin period $P_{spin}$. A variation in the assumption of  $r_{in} = r_{M} = r_{CO}$ would therefore result in variation of $P_{spin}$ and $\nu_{QPO}$ pairs for corresponding $B$ and $\dot{M}$ pairs.

\begin{figure*}
	\includegraphics[width=\textwidth,trim={2cm 2cm 2cm 2cm},clip]{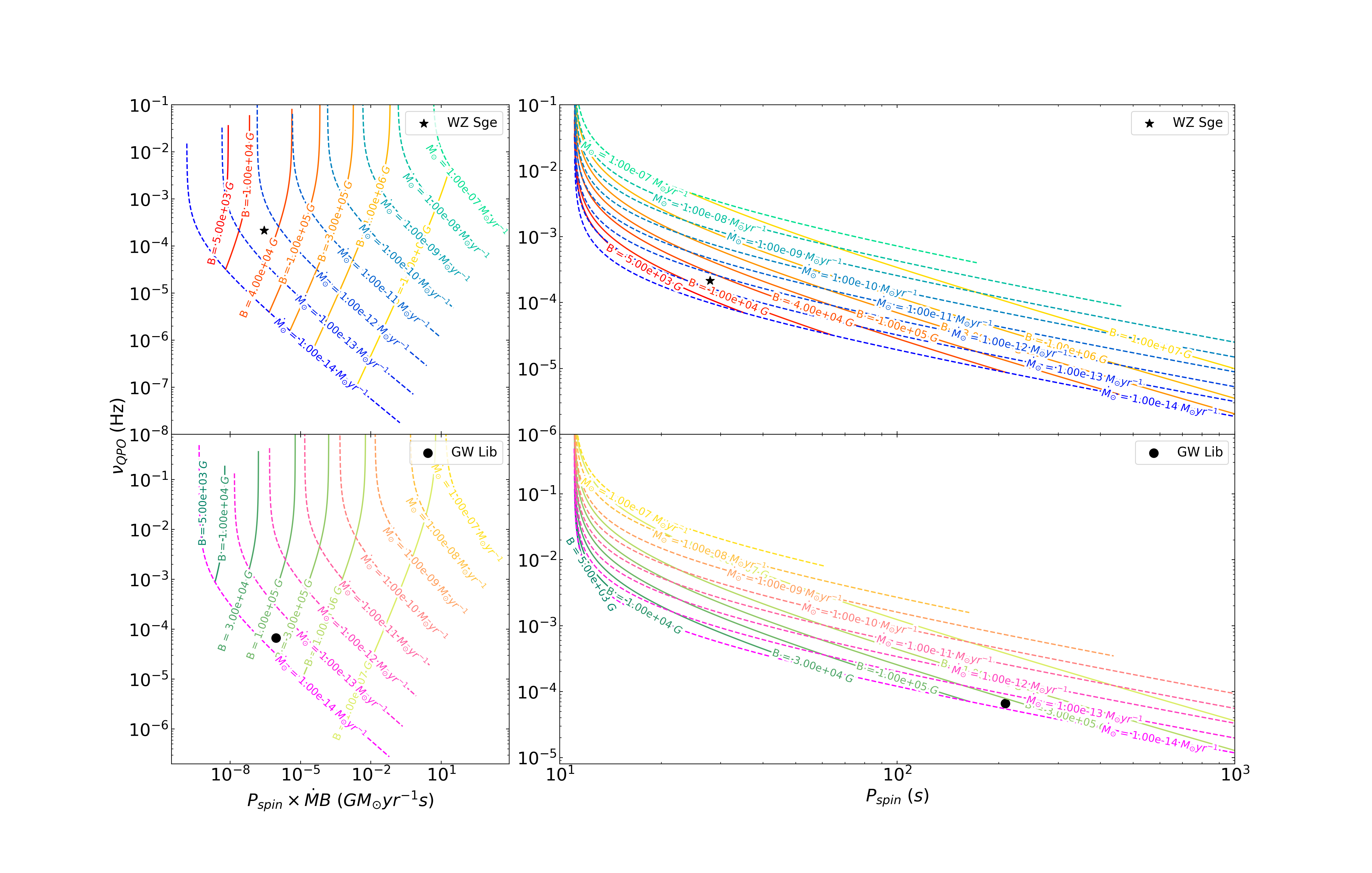}
    \caption{\textit{Right}: Model QPO frequency as a function of the corresponding accretor spin. \textit{Left}:Model QPO frequency as a function of the corresponding accretor spin scaled by the accretion rate and magnetic field strength of the accretor. In all panels the dashed lines denote lines of equal $B$ field and solid lines denote lines of equal accretion rate. \textit{Top panels}: Fiducial parameters corresponding to WZ Sge from Table \ref{tab:model_fid}. The star corresponds to the observed QPO and spin, scaled by $B$ and accretion rate on the right. \textit{Bottom panels}: Same as top panels but with fiducial parameters of GW Lib from Table \ref{tab:model_fid}. The filled circle represents the observed QPO and scaled spin of GW Lib.}
    \label{fig:QPOmodel}
\end{figure*}

Figure \ref{fig:QPOmodel} demonstrates the dependence of the model on these 4 parameters. In the right hand side panels the QPO frequency is plotted as a function of $P_{spin}$. The dashed lines then show lines of equal $B$ and the solid lines the lines of equal $\dot{M}$. A QPO measurement can in such a case be used to put constraints on $B$ and and accretion rate for a given spin. Since there is a strong correlation between the $B$ and accretion rate in the model, the same plots are also shown in the left hand side panels where $P_{spin}$ is scaled by the magnetic field strength and the accretion rate. 

\begin{table}
	\centering
	\caption{Fiducial model parameters for the magnetically driven precession model for QPOs for AWDs and NS XRBs. For the main parameters that are varied through the model, the magnetic field of the accretor $B$ and the accretion rate $\dot{M}$, the ranges are cited instead. For AWDs, 2 values of viscosity and $\eta$ are given to represent WZ Sge and GW Lib like objects.}
	\label{tab:model_fid}
	\begin{tabular}{lcr} 
		\hline
		Model Parameter & AWDs & NS XRBs \\
		\hline
		M ($M_{\odot}$) & 0.8 & 1.4 \\
        R  & 0.01 $R_{\odot}$ & 10 km \\
        B (G) & 10$^{3}$ $-$ 10$^{8}$ & 10$^{7}$ $-$ 10$^{10}$  \\
        $\dot{M}$ ($M_{\odot}yr^{-1}$) & 10$^{-12}$ $-$ 10$^{-7}$  & 10$^{-12}$ $-$ 10$^{-7}$  \\
        $\alpha$ & 0.006 (WZ Sge), 0.003 (GW Lib) & 0.006 \\
        $\eta$ & 0.1 (WZ Sge), 0.05 (GW Lib) & 0.1 \\
        $\theta$ ($^{\circ}$) & 45 & 45 \\
        A & 0.6 & 0.6 \\
		\hline
	\end{tabular}
\end{table}

By evaluating the magnetically driven precession model for the fiducial model values and the $B$ and $\dot{M}$ ranges listed in Table \ref{tab:model_fid} it is possible to estimate these parameters and the corresponding spin period of the accretor. The WD mass is fixed to 0.8 $M_{WD}$ for all objects as a good approximation for the mean of the AWD accretor mass distribution \citet{Zorotovic2011A&A...536A..42Z,Pala2022MNRAS.510.6110P}. The WD radius is then derived from \citet{Hamada1961ApJ...134..683H,Nauenberg1972ApJ...175..417N} using the WD mass. $\eta$, the parameter relating Alfv\' en and magnetospheric radius, is set to 0.1 and 0.05. For strongly magnetic systems such as NS and magnetic AWDs, it is by convention set to 0.5 \citep{Ghosh1979ApJ...234..296G,Monkkonen2022MNRAS.515..571M}. However, 
the model requires the magnetospheric radius to be equal to the co-rotation radius. This is not likely to always be the case and a smaller value of $\eta$ required here may reflect that. $\theta$ is set to the midpoint value of 45$^{\circ}$. $A$ is fixed to 0.6 as this is the average of the range determined for the parameter in the framework of this model in \citet{Shirakawa2002a}.

To compare how successfully the model predicts $B$, $\dot{M}$, and spin for a particular QPO frequency, it is possible to use the cases of WZ Sge and GW Lib. To estimate $B$ and $\dot{M}$ from the model a fine grid of $B$ and $\dot{M}$ is computed with a pair of spin and QPO frequency corresponding to each pair of $B$ and $\dot{M}$. Then for a given measurement of spin and QPO frequency with their corresponding uncertainties the corresponding range of $B$ and $\dot{M}$ can be extracted.

WZ Sge has an estimated spin period of 27.87$\pm$0.01\,s as measured by \citet{Patterson1980ApJ...241..235P}. Later in \citet{Patterson1998PASP..110..403P} the magnetic field is invoked to be between 1 $-$ 5 $\times$ 10$^{4}$ G. This is necessary to explain the disappearance of the potential spin frequency during an outburst when the accretion rate of a dwarf nova such as WZ Sge rises by a factor of $\sim$1000 from the quiescent rate of $\sim$1 $\times$ 10$^{-11}$ $M_{\odot}yr^{-1}$. With this information it is possible to compare the observables of WZ Sge to the parameter space in the top panel of Figure \ref{fig:QPOmodel}. The QPO and spin frequency with the assumed $B$ range and $\dot{M}$ correspond to the correct range within the parameter space, showing that the model can reproduce reasonable results for the fiducial values of fixed parameters, with B = 2.5$\pm$0.1$\times$10$^{4}$ G and $\dot{M}$ = 4.0$\pm$0.3$\times$10$^{-13}$ $M_{\odot}yr^{-1}$. This value is at odds with the previously reported quiescent accretion rate for WZ Sge. An important point to consider is that in the model the accretion rate represents the accretion rate onto the accretor, not the mass transfer rate from the donor. Furthermore, because there are a set number of fixed fiducial parameters, an error estimate on the model's parameter prediction is only meaningful when considering the potential variability of the free parameters that remain fixed. Hence the errors here only represent a snapshot of the real uncertainty contours and so should be treated as a lower limit on the real errors.

Similarly for GW Lib, the spin period was measured to be 209\,s from the UV line widths \citep{Szkody2012ApJ...753..158S}. As the recurrence timescale for GW Lib is very long ($\sim$ years) the quiescent viscosity is also low \citep{Howell1995ApJ...439..337H}. Adapting the fiducial model parameters from Table \ref{tab:model_fid} the resulting $B$ field and accretion rate estimate are B = 2.0$\pm0.1\times$10$^{5}$ G and $\dot{M}$ = 2.2$\pm$0.3$\times$10$^{-14}$ $M_{\odot}yr^{-1}$. This leads to a higher magnetic field compared to WZ Sge and extremely low accretion rate. Nevertheless, such a low accretion rate has been reported for quiescent state of GW Lib by \citet{Hilton2007AJ....134.1503H}. However given the lack of constraint on some of the model parameters, this should be treated more as a rough estimate. The parameter space for GW Lib and its adapted fiducial parameters are shown in the bottom panel of Figure \ref{fig:QPOmodel}.

\begin{table}
	\centering
	\caption{Estimates of magnetic field strength, accretion rate and spin period from the magnetically driven precession model of QPO in AWDs. the values should be treated as possible ranges and not as best fit values. In the case of WZ Sge and GW Lib the parameters can be constrained better using the observational estimates of $P_{spin}$, which are noted in the table instead and denoted by $^{*}$. For objects with unknown spin, an estimate of accretion rate is used so that $^{a}$ \citet{Veresvarska2024}, $^{b}$ \citet{Godon2018ApJ...862...89G} and $^{c}$ $\sim 10^{-11} - 10^{-10}$ $M_{\odot}yr^{-1}$ as a representative value for the WZ Sge-type Dwarf Nova V3101 Cyg. CP Pup and T Pyx use the fiducial parameters of WZ Sge and V3101 Cyg of GW Lib.}
	\label{tab:model_B_Mdot}
	\begin{tabular}{lccr} 
		\hline
		Object & B (G) & $\dot{M}$ $(M_{\odot}yr^{-1})$ & $P_{spin}$ (s) \\
		\hline
		CP Pup & $2-3\times10^{5}$ & $1-2\times10^{-10}$ $^{a}$ & $72-84$ \\
        WZ Sge & 2.5$^{+0.1}_{-0.1}\times10^{4}$ & 4.0$^{+0.3}_{-0.3}\times10^{-13}$ & 27.87$^{+0.01}_{-0.01}$ $^{*}$ \\
        GW Lib & 1.97$^{+0.14}_{-0.12}\times10^{5}$ & 2.2$^{+0.3}_{-0.3}\times10^{-14}$ & 209$^{*}$ \\
        T Pyx & $9\times10^{6} - 1\times10^{7}$ & $10^{-7} - 10^{-6}$ $^{b}$& $148-345$ \\
        V3101 Cyg & $3\times10^{3} - 1.2\times 10^{4}$ & $10^{-11} - 10^{-10}$ $^{c}$& $99-165$ \\
		\hline
	\end{tabular}
\end{table}

As there are no estimates for spin period, $B$ or $\dot{M}$ for any of the other systems, only very broad ranges can be provided for the model parameters, adopting $\alpha=$0.006 and $\eta=$0.1. The only exception is V3101 Cyg, which has theoretically predicted $\alpha < 0.005$ \citep{Hameury2021A&A...650A.114H} and so the fiducial parameters of GW Lib are used instead. In such a case it is impossible to provide reasonable constraint on any of the parameters. However, previously reported values of accretion rate can help constrain the estimates of $B$ field and spin. The accretion rates used are $1-2 \times 10^{-10}$ $M_{\odot}yr^{-1}$ from \citet{Veresvarska2024} for CP Pup, $\sim10^{-7} - 10^{-6}$ $M_{\odot}yr^{-1}$ for T Pyx from \citet{Godon2018ApJ...862...89G} and $\sim10^{-11} - 10^{-10}$ $M_{\odot}yr^{-1}$ for V3101 Cyg which is a WZ Sge type Dwarf Nova. The corresponding $B$ and spins are listed in Table \ref{tab:model_B_Mdot}, alongside the accretion rate estimates for WZ Sge and GW Lib. The obtained $B$ field is below the standard values of confirmed magnetic systems, apart from T Pyx, and all estimated spin ranges are under 3 minutes, apart from T Pyx, where the range of possible spins reaches $\sim$6 minutes.

It is however necessary to note that all such deductions on the magnetic field strength and accretion rate are highly dependent on a good understanding of the system. As of now, there are no known systems in which there is sufficient certainty to constrain the values of the fiducial parameters. As a consequence all deduced constraints on $B$ and $\dot{M}$ are simply a reflection of the constraints on the spin and QPO frequency measurements. Hence the results reported in Table \ref{tab:model_B_Mdot} should be treated with extreme caution and mostly serve as a prediction of the potential precision, subject to future precise measurements of parameters in Table \ref{tab:model_fid}.

\subsubsection{QPO model implications for spin, B field and accretion rate measurement}
\label{sss:QPO_model_disc}

Figure \ref{fig:QPOmodel} shows that for the current set up of the model, all AWDs showing QPOs with similar characteristics as WZ Sge would be expected to have a relatively fast spin $\sim$20 $-$ 200\,s WD. This is assuming a low magnetic field $B \lesssim 10^{6}$ G and low accretion rate $< 10^{-10}$ $M_{\odot}yr^{-1}$ and a similar frequency range for the QPOs as observed in the systems here. The extremely low accretion rates recovered by the model here should be taken as representative of the instantaneous accretion rate onto the accretor, not the mass transfer rate from the donor star. Hence such low values may not be unrealistic for the accumulation of mass between Dwarf Novae outbursts in WZ Sge and GW Lib, in particular in the case of GW Lib \citep{Hilton2007AJ....134.1503H}.

For slowly spinning systems $\gtrsim$ 1000~s, the QPOs in weakly magnetised systems adopting parameters self-similar to the fiducial ones used here would yield QPO frequencies $\lesssim$ 10$^{-6}$ Hz (278 hrs). This would provide a considerable observational challenge with the current data. New potential missions such as \plato\ may help uncover these in the future. For magnetic systems, monitoring of close-by and bright magnetic systems, such as the Intermediate Polars V1223 Sgr or FO Aqr, may unveil new QPOs in systems with alternative values of $B$ and direct spin observations to use for model verification. These potential detections could then be used to put a better constraint on some of the fiducial parameters in Table \ref{tab:model_fid}.

Given that the QPOs reported here are mostly observed in either dwarf novae or nova-likes the potential variability of the QPO with a large change of accretion rate could also provide important insight on the nature of the QPOs themselves and the systems in which they occur as well as their driving mechanism.

\subsection{Break and QPO correlation in XRBs and AWDs and QPO driving mechanism implications}
\label{ss:QPO_break_disc}

\begin{figure*}
	\includegraphics[width=\textwidth]{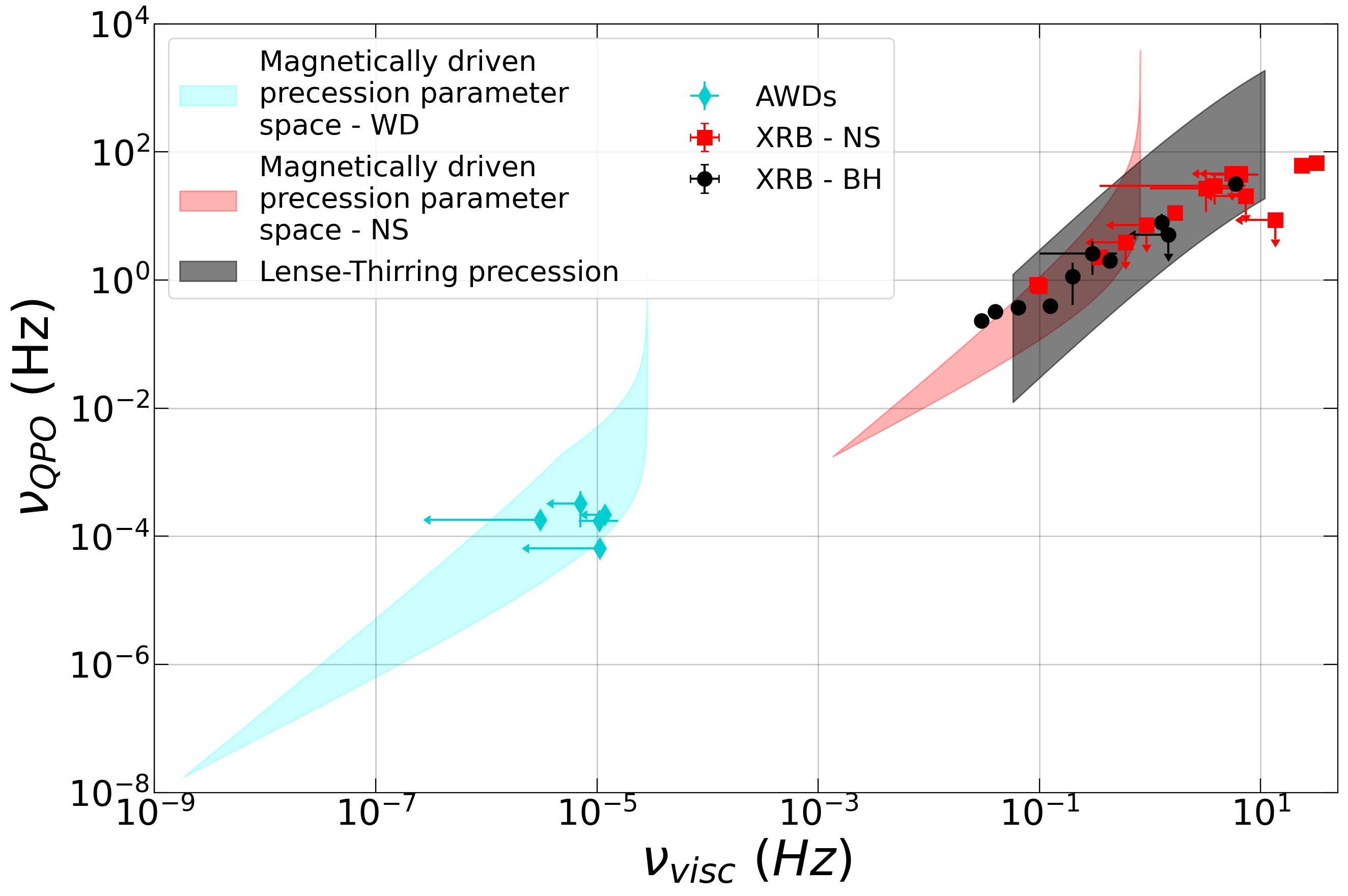}
    \caption{QPO frequency as a function of the viscous frequency at the QPO radius. The shaded regions correspond to the predicted model parameter space of QPOs and their viscous frequency. The lowest shaded region between $\nu_{visc}\sim10^{-10}-10^{-4}$ Hz represents the model parameter space of AWD QPOs as predicted by magnetically driven precession. For NS XRBs the magnetically driven precession model predicts a parameter space shown by the shaded region between $\nu_{visc} \sim 10^{-4} - 10^{0}$ Hz. The partially overlapping shaded region between $\nu_{visc} \sim 10^{-1} - 10^{2}$ Hz represents the parameter space of QPOs in BH XRBs as predicted by Lense-Thirring precession. The observed QPOs with their corresponding low frequency breaks ($\nu_{visc}$ are denoted for AWDs in diamonds, for NS XRBs in squares and in filled circles for BH XRBs.) }
    \label{fig:QPOsXRBWD}
\end{figure*}

Type-C and HBO QPOs in XRBs bear several observational similarities to the QPOs reported here in AWDs. An important characteristic of these QPOs in XRBs is their correlation to a low-frequency break \citep{Wijnands1999}. The QPOs in AWDs reported here appear to follow the same correlation. Type-C and HBO QPOs in XRBs are usually explained by the relativistic effect of frame dragging. In such a case it can be assumed that the PSD break is associated to the viscous frequency at the inner-edge of the precessing flow \citep{Ingram2009MNRAS.397L.101I,Ingram2011MNRAS.415.2323I}.

This cannot be the case in the non-relativistic AWDs. However in the framework of the magnetically driven precession model as described in Section \ref{ss:QPOmodel_disc} the observational characteristics of the QPOs are similar, despite a different driver behind the behaviour. A precessing inner flow of the accretion disc would also in this case be expected to produce a PSD break at the viscous frequency associated to the corresponding radius. This would have the effect of producing a break at a lower frequency than the QPO and a relation between the two quantities, as observed in Figure \ref{fig:QPOsXRBWD}. 

In Figure \ref{fig:QPOsXRBWD} we reproduce the data from Figure \ref{fig:QPOsbreakXRBWD}, but now overplay several models. As in Figure \ref{fig:QPOsbreakXRBWD}, the black circles represent the BH QPOs and their breaks from \citet{Wijnands1999}, the red squares represent the NS QPOs and their corresponding breaks from \citet{Wijnands1999}. The cyan diamonds represent the AWD QPOs and breaks from Table \ref{tab:PSDfit}. The shaded regions represent the model parameter spaces for Lense-Thirring precession for BH XRBs, magnetically driven precession for NS XRBs and AWDs. The cyan shaded regions overlapping with the observed AWD QPOs represents the parameter space in which QPOs in AWDs can be explained by magnetically driven precession for all parameters shown in Table \ref{tab:model_fid}. The viscous frequency, $\nu_{visc}$, was obtained from the dynamical frequency, $\nu_{dyn}$ at the QPO radius such that $\nu_{visc} = \alpha (\frac{H}{R})^{2} \nu_{dyn}$, and by assuming a value for $\alpha \left( \frac{H}{R} \right)^{2} \sim 5 \times 10^{-5}$. $\alpha \left( \frac{H}{R} \right)^{2}$ is treated here as a standalone parameter, not a combination of viscosity $\alpha$ and the scale height ratio $\frac{H}{R}$. This is because in the context of AWDs there have been measurements of $\alpha \left( \frac{H}{R} \right)^{2}$ parameters that can be used to estimate this parameter \citep{Scaringi2014,Veresvarska2023}. Whereas estimates of $\alpha$ have been measured \citep{Howell1995ApJ...439..337H,Kotko2012}, no tangible constraint is known for $\frac{H}{R}$. Therefore treating $\alpha \left( \frac{H}{R} \right)^{2}$ as a single parameter based on previous measurements is deemed as the more appropriate assumption. The value is chosen as the middle of the expected range for this parameter for AWDs in quiescence. The potential range is considered for $\alpha$ to be between 10$^{-3}$ for long recurrence dwarf novae such as GW Lib \citep{Howell1995ApJ...439..337H}  up to 10$^{-1}$ for high accretion rate systems \citep{Kotko2012}, such as T Pyx. The estimated range for $\alpha \left( \frac{H}{R} \right)^{2} \sim 1 \times 10^{-7} - 1 \times 10^{-3}$. The implications of the uniform assumption of $\alpha \left( \frac{H}{R} \right)^{2} \sim 5 \times 10^{-5}$ with the range of $\alpha \sim 1\times10^{-3} - 0.1$ \citep{Howell1995ApJ...439..337H,Kotko2012} is that the $\frac{H}{R} \sim 0.2 - 0.02$.

The x-axis of Figure \ref{fig:QPOsXRBWD} is the viscous frequency at the QPO radius. This demonstrates that all the QPOs reported here could be explained by the magnetically driven precession model, bearing in mind the upper limit on the break frequencies in all objects apart from WZ Sge. Therefore a potential test of this model and correlation would be the measurement of break frequencies for these systems and whether it deviates from the current limit by more than an order of magnitude. Such a test would however require more data, either an extension of the existing \tess\ time series or a new long-term mission such as \plato\,.

The red shaded region overlapping with observed QPOs in XRBs and ranging from $\sim$ 10$^{-4}$ $-$ 10$^{0}$ Hz for the viscous frequency corresponds to the parameter space where the NS XRB QPOs and their PSD breaks can be explained by the magnetically driven precession. The parameter space was constructed for a NS with $M_{NS} \sim 2 M_{\odot}$, $R_{NS} \sim 10$km,  $B$ $\sim 10^{7}$ $-$ $10^{10}$ G, the same accretion rate range, $\theta$ and $A$ as in Table \ref{tab:model_fid}. $\eta$ was fixed to 0.1 and $\alpha$ to 0.006 for consistency with the AWDs. $\alpha \left( \frac{H}{R} \right)^{2}$ was assumed the same as in AWDs. 

The black shaded region overlapping with QPOs in XRBs from $\sim$ 10$^{-1}$ $-$ 10$^{2}$ Hz corresponds to the parameter space where the QPOs in XRBs can be explained by Lense-Thirring precession. QPOs driven by Lense-Thirring precession occur at a characteristic frequency which is related to the precession of a solid disc as detailed in \citet{Ingram2009MNRAS.397L.101I} :

\begin{equation}
    \nu_{LT} = \frac{\left(5 - 2 \zeta\right)}{\pi \left(1 + 2\zeta\right)} \frac{a \left(1 - \left(\frac{r_{i}}{r_{o}}\right)^{\frac{1}{2}+\zeta}\right)}{r_{o}^{\frac{5}{2} - \zeta} r_{i}^{\frac{1}{2}+\zeta} \left[ 1 - \left( \frac{r_{i}}{r_{o}} \right)^{\frac{5}{2} - \zeta}\right]} \frac{c}{R_{g}}
\end{equation}

\noindent where $a$ is the BH spin, $c$ is the speed of light in vacuum, $r_{i}$ the inner radius of the precessing flow in the units of $R_{g}$ and $r_{o}$ the outer radius of the precessing flow in the units of $R_{g}$. $R_{g}$ represents the gravitational radius $R_{g} = \frac{GM}{c^{2}}$. $\zeta$ is the index governing the radial dependence of surface density and is here fixed to 0 as from simulations as in \citet{Ingram2009MNRAS.397L.101I}. The mass range accounted for in the parameter space spans BH mass from 3 $-$ 20 $M_{\odot}$ and spins from 0 to 0.998. The outer radius of the precessing region is considered between 1.01 $-$ 50 $R_{g}$ and associated to $r_{o}$. The inner radius is here fixed to the innermost stable circular orbit (ISCO) based on the spin. This can predict larger QPO frequencies in black holes than are observed and the upper limit of the parameter space should be treated with caution. This is usually accounted for by setting the inner disc radius to the bending wave radius \citep{Ingram2009MNRAS.397L.101I} which would require the assumption of $\frac{h}{r}$ ratio. As this is unknown and could introduce another free parameter, the inner disc radius is here fixed at ISCO for simplicity. The viscous frequency is determined for radius $r_{o}$ with $\alpha \left( \frac{H}{R} \right)^{2} \sim 10^{-3}$ as is usually assumed for BH XRBs \citep{Ingram2009MNRAS.397L.101I}. 

For all the parameter spaces detailed above for Lense-Thirring precession and magnetically driven precession it is clear that that both follow the same observational trend behind the QPO and break relation, irrespective of the driver mechanism of the QPO. This unfortunately also means that the relation cannot be used to disentangle the nature of the driving mechanism for QPOs in XRBs since the 2 parameter spaces overlap. However, the non-relativistic AWDs show tentatively, whilst bearing in mind the uncertainty from the upper limits on most breaks in AWDs, that magnetically driven precession is a viable method for inducing QPOs in weakly magnetised AWDs.

\section{Conclusions}
\label{s:Conclusions}

QPOs are a well known characteristic in both XRBs and AWDs. In XRBs they are a well defined phenomena with different types according to their observational characteristics and correlation with broad-band structure. However, in AWDs, QPOs usually refer to unexplained, transitional and somewhat coherent signals observed in the light curves. Here we report 5 persistent QPOs in low magnetic field AWDs showing similar properties to type-C QPOs in XRBs, such as harmonics. A tentative link to the QPOs in XRBs is that the QPOs in AWDs seem to follow the correlation of XRB QPOs with a broad band low frequency break, when assuming that the upper limits on the break frequencies are representative of their true value. The break is usually associated with the viscous frequency of the outer part of the precessing flow of the disc where the QPOs are generated. Furthermore, the reported QPOs in AWDs show a low frequency broad-band structure, which can be fitted with a zero-centred Lorentzian as in XRBs or with a power-law, at a frequency where such a flow transition would be expected.

Due to the non-relativistic nature of AWD systems, we propose that the QPOs are driven by a weak magnetic field of the accretor being misaligned to the disc normal and causing precession of the inner part of the accretion disc. This model was previously considered for being part of the driving mechanism behind the low frequency QPOs in NS XRBs and potentially T Tauri stars. Here we show an implementation for AWDs, whose magnetic fields lie in between those of NS and T Tauri stars. 

WZ Sge shows a QPO at $\sim$ 2 $\times$ 10$^{-4}$ Hz ($\sim$77 min) and GW Lib at $\sim$ 7 $\times$ 10$^{-5}$ Hz ($\sim$4.3 hrs), with both exhibiting harmonics. They are also the only systems in the sample with spin period measurements. We report that the model prediction for a QPO at these frequencies requires accretion rate and spin that are consistent with the observationally reported values. This could present a possibility for magnetic field and spin estimates in weakly magnetic AWDs where direct measurements of these quantities are not yet possible. As an example we give estimates for these parameters for the rest of our sample. However, such predictions, together with the corresponding uncertainties, can only be obtained within the framework of the inherent assumptions of the model, once reasonable estimates on the fiducial parameters can be obtained. Therefore the results here can only serve as an illustration of the potential capabilities of the model, until a more detailed measurement of the fiducial parameters is obtained with the help of future missions such as \plato\,.

\section*{Acknowledgements}

This paper includes data collected with the TESS mission, obtained from the MAST data archive at the Space Telescope Science Institute (STScI). Funding for the TESS mission is provided by the NASA Explorer Program. STScI is operated by the Association of Universities for Research in Astronomy, Inc., under NASA contract NAS 5–26555. This publication makes use of data products from the Wide-field Infrared Survey Explorer, which is a joint project of the University of California, Los Angeles, and the Jet Propulsion Laboratory/California Institute of Technology, funded by the National Aeronautics and Space Administration. MV acknowledges the support of the Science and Technology Facilities Council (STFC) studentship ST/W507428/1. SS is supported by STFC grant ST/T000244/1 and ST/X001075/1. DdM acknowledges financial support from ASI-INAF agreement and INAF Astrofund-2022 grants. AI acknowledges support from the Royal Society. ZAI acknowledges the support of the Science and Technology Facilities Council (STFC) studentship ST/X508767/1.

\section*{Data Availability}

The \tess\ data used in the analysis of this work is available on the MAST webpage \url{https://mast.stsci.edu/portal/Mashup/Clients/Mast/Portal.html}. The \asassn\ data \citep{Shappee2014,Kochanek_2017} used for the calibration of \tess\ data is available on the \asassn\ webpage \url{https://asas-sn.osu.edu/}.



\bibliographystyle{mnras}
\bibliography{refs} 




\appendix

\section{\tess\ light curves}
\label{a:LC}

\begin{figure*}
	\includegraphics[width=\textwidth]{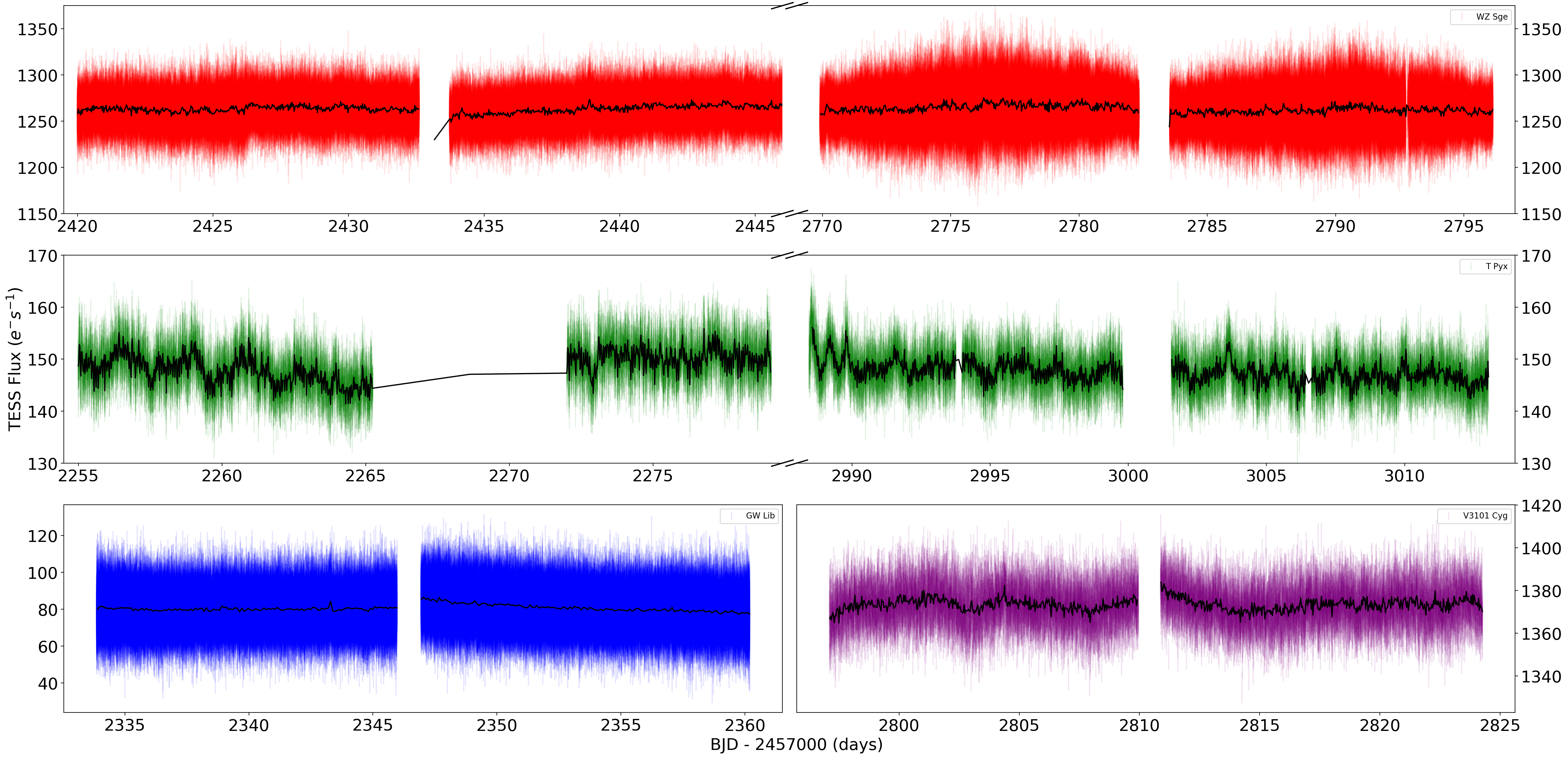}
    \caption{Light curve of WZ Sge, GW Lib, T Pyx and V3101 Cyg showing all sectors of \tess\ data as described in Table \ref{tab:obs} (CP Pup is shown in Figure 1 of \citealt{Veresvarska2023}). The gap between the individual \tess\ sectors in WZ Sge and GW Lib has been excluded. The solid line in each panel denotes the running average of the light curve on the QPO timescale.}
    \label{fig:LC}
\end{figure*}

Figure \ref{fig:LC} shows the \tess\ light curves of AWDs used in this work apart from CP Pup as noted in Table \ref{tab:obs}. To demonstrate more clearly the variability on the QPO timescale a running average of the light curve is overlaid on top of the data in solid black line.

\section{Linear PSDs of QPOs}
\label{a:PSD}

\begin{figure*}
	\includegraphics[width=\textwidth]{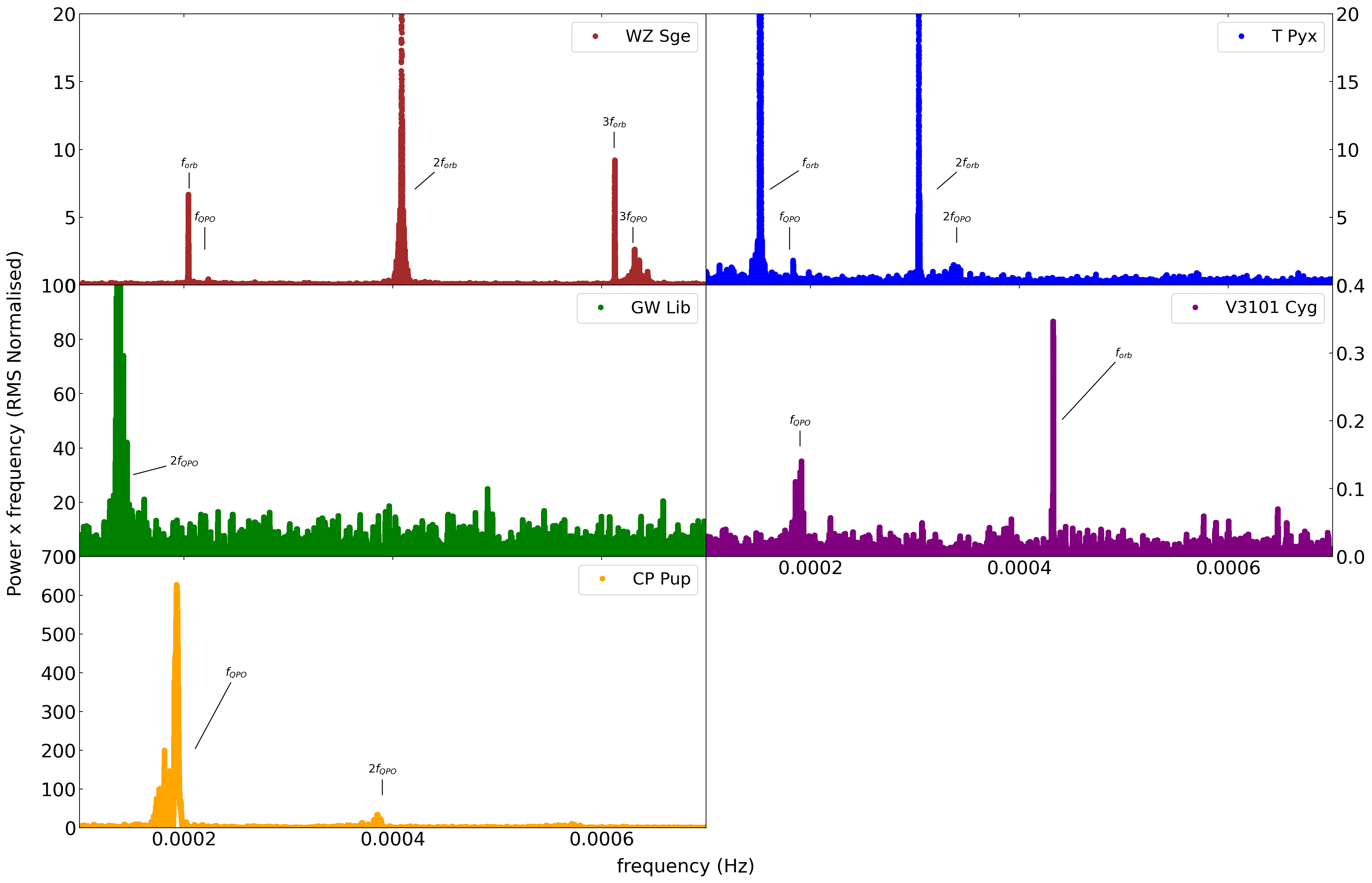}
    \caption{A zoom-in on a section of PSD of light curves from Table \ref{tab:obs} with labeled QPO and orbital period signals. The frequency range is fixed to be the same for all panels. Some of WZ Sge harmonics were excluded for better visualisation.}
    \label{fig:PSD_lin}
\end{figure*}

Figure \ref{fig:PSD_lin} shows the linear non-averaged PSD of the light curves from Table \ref{tab:obs}. The PSDs are zoomed in on the frequency range where the QPOs and their harmonics are the most dominant. In this case the orbital period signals are not removed to demonstrate the difference between the QPOs and a coherent orbital period. For WZ Sge the first harmonic of the QPO is not visible due to the first harmonic of the orbit. Other high frequency harmonics of the QPO and orbit are present but not shown in the frequency range displayed here. For GW Lib only the first harmonic of the QPO is shown as the fundamental of the signal observed in \citet{Chote2021MNRAS.502..581C}.

\section{QPO Model: Magnetically Driven Precession Model}
\label{a:model}

The magnetically driven precession model of QPOs discussed in this Section was developped by \citet{Lai1999} for the purpose of explaining QPOs in NSs and T Tauri stars. In this frame work the accretor's spin axis is misaligned to the angular momentum of the accretion disc causing magnetic and warping torques. These torques warp the inner part of the disc which then precesses around.

The magnetospheric radius at which the accretion flow is disrupted by the accretor's magnetic field B is given by

\begin{equation}
    r_{M} = \eta \left( \frac{2 \pi^{2}}{\mu_{0}^{2}} \frac{\mu^{4}}{G M \dot{M}^{2}} \right) ^{\frac{1}{7}},
    \label{eq:Rm}
\end{equation}

where $\eta$ is a dimensionless parameter describing the geometry of the accretion flow and the relation between the Alfv\' en and magnetospheric radius. It is usually set to 0.5 for magnetic systems, but a lower value could indicate that the assumption of $r_{M} = r_{CO}$, typically assumed for magnetic systems, is not valid (i.e. for a low magnetic field). Such a scenario (low value of $\eta$) could also be possible for a system with high inclination. $\mu$ is the stellar magnetic dipole  moment such that $\mu = B R^{3}$, $M$ the accretor mass, $R$ the accretor radius, $\dot{M}$ its accretion rate, G the gravitational constant and $\mu_{0}$ the vacuum permeability. Following \citet{Lai1999,Pfeiffer2004}, the magnetospheric radius from \ref{eq:Rm} is set to be equal to the co-rotation radius:

\begin{equation}
    r_{CO} = \left( \frac{GMP_{spin}^{2}}{4 \pi^{2}} \right) ^{\frac{1}{3}}
    \label{eq:Rc},
\end{equation}

where $P_{spin}$ is the spin period of the accretor. This assumption of co-rotation radius being equal to magnetospheric radius is a simplification and may not be entirely realistic. This radius is also set to be equal to the inner disc radius $r_{in}$. With these assumptions, the global precession frequency of the inner flow is given by

\begin{equation}
    \nu_{QPO} = A \frac{\Omega_{p} \left(r_{in} \right)}{2 \pi} = \frac{A}{2 \pi} \frac{\mu^{2}}{\pi^{2} r^{7} \Omega \left(r \right) \Sigma \left( r \right) D \left( r \right)} F \left( \theta \right),
    \label{eq:nuQPOa}
\end{equation}

where $A$ is related to the offset between the global precession frequency and the magnetically driven precession frequency. Depending on the details of the disc structure at $r_{in} = r_{M}$, $A \simeq 0.3 - 0.85$ as determined by \citet{Shirakawa2002b}. $\Omega \left(r \right)$ is the Keplerian frequency at a radius $r$, $\Sigma \left( r \right)$ is the surface density as determined in Equation 5.41 in \citet{Frank2002}. $D \left( r \right)$ is a dimensionless function given by

\begin{equation}
    D \left( r \right) = max \left( \sqrt{\frac{r^{2}}{r_{in}^{2}} - 1} , \sqrt{\frac{2 H \left( r \right)}{r_{in}}} \right),
    \label{eq:Dr}
\end{equation}

where $H \left( r \right)$ is the half-height of the disc at radius $r$, as defined in Equation 5.41 in \citet{Frank2002}. $F \left( \theta \right)$ is a function that depends on the dielectric property of the disc, so that $F \left( \theta \right) = 2 f cos^{2} \theta - sin^{2} \theta$, where $\theta$ is the angle between the magnetic moment of the accretor and the angular momentum of the disc. $f$ is a dimensionless number between 0 and 1, so that for $f=1$ all the vertical field is screened out by the disc, whereas for $f=0$ only the spin-variable vertical field is. We follow  \citet{Pfeiffer2004} in taking $f=0$ throughout this work for simplicity.

Though the QPO can happen at any radius $r$ the degeneracy this would create would not allow to test the model through observations. However, assuming $r_{in} = r_{M} = r_{CO} = r$ mitigates this. Through $r_{in} = r_{M} = r_{CO}$, the precession and QPO frequency depend on the spin of the accretor, providing a potential handle on the spin period of systems in which it cannot be directly measured, such as weakly magnetic AWDs. Naturally the model is dependent on many other parameters, chiefly among them the strength of the intrinsic magnetic field strength of the accretor $B$, the accretion rate of the system $\dot{M}$ and the viscosity of the disc $\alpha$.


\bsp	
\label{lastpage}
\end{document}